\documentclass[reqno,12pt]{article} 
 \usepackage{amssymb,amsmath,amscd} 

\usepackage{color}
\definecolor{blue}{rgb}{0.00,0.00,1.00}
\definecolor{red}{rgb}{1.00,0.00,0.63}
\newcommand{\blue}{\color{blue}}

\usepackage[a4paper,left=30mm,textwidth=148mm,headsep=4mm,top=25mm,bottom=25mm]{geometry}
\usepackage{hyperref}

  \renewcommand\baselinestretch{1.2}

   \def\({\!\left(}
   \def\){\right)\!}

   \def  \dx    {\partial_x}
   \def  \dt    {\partial_t}

   \def  \div   {\nabla\!\cdot}

    \def \R   {\mathbb{R}}

    \def \Td   {\nabla}
    \def \Lp   {\Delta}

    \def \n {{\ifmmode{\msyb N}\else$\msyb N$\fi}}
    \def \q {{\ifmmode{\msyb Q}\else$\msyb Q$\fi}}
    \def \r {{\ifmmode{\msyb R}\else$\msyb R$\fi}}
    \def \c {{\ifmmode{\msyb C}\else$\msyb C$\fi}}

 \def\be#1\ee{\begin{equation}#1\end{equation}}
 \def\bay#1\eay{\!\!\!\left\{\!\!\begin{array}{l}%
      #1\displaystyle\end{array}\right.}
 \def\bln#1\eln{\begin{aligned}#1\end{aligned}}

 \def\bma#1\ema{{\allowdisplaybreaks\begin{align}#1\end{align}}}
 \def\nnm{\notag}

 \def\bgr#1\egr{{\allowdisplaybreaks\begin{gather}#1\end{gather}}}
 \def\bman#1\eman{\begin{eqnarray*}#1\end{eqnarray*} }
 
 \def\ef#1{(\ref{#1})}
 
 \def\qef#1{$(\ref{#1})$}
 
       \def\no\noindent
       \let\bf=\textbf

       \theoremstyle{plain}
       \newtheorem{lemma}{\textbf{Lemma}}[section]
       \newtheorem{theorem}[lemma]{\textbf{Theorem}}
       \newtheorem{proposition}[lemma]{\textbf{Proposition}}

       \newtheorem{remark}{\textbf{Remark}}

       \def\oplem#1{\begin{lemma}\, {\rm #1}\, \it }
       \def\cllem{\end{lemma}\rm \par }

       \def\opthm#1{\begin{theorem}\, {\rm #1}\, \it }
       \def\clthm{\end{theorem}\rm \par }

       \def\opprop{\begin{proposition} }
       \def\clprop{\end{proposition}\rm \par }

       \def\opdef{\begin{definition}}
       \def\cldef{\end{definition}\rm\par }

       \def\oprem{\begin{remark}\rm}
       \def\clrem{\end{remark}\rm \par}

       \def\demo{{\bf Proof: \ }\rm }
       
\begin{document}

%
\title{\Large\textsc{Optimal decay rate of the
compressible Navier-Stokes-Poisson system in $\R^3$ }}
\author{{Hai-Liang Li$^1$, A. Matsumura$^2$, Guojing Zhang$^3$}
\\[3mm]
{\normalsize\it $^{1}$Department of Mathematics and
 Institute of Mathematics and Interdisciplinary } \\
{\normalsize\it Science, Capital Normal University, P.R.China}\\
{\normalsize\it  $^{2}$Graduate School of Information Science and
                Technology, Osaka University, Japan} \\
{\normalsize\it  $^{3}$Department of Mathematics,
     Harbin University of Technology, P.R.China}\\
 {\small\it  e-mail:\, hailiang.li.math@gmail.com, \quad
  akitaka@math.sci.osaka-u.ac.jp}\\
 {\small\it  zhanggj112@yahoo.cn}
  }

\date{} 
 \maketitle

\pagestyle{myheadings}
\markboth{Compressible Navier-Stokes-Poisson}%
{\small Li, Matsumura, $\&$ Zhang}


%
\begin{abstract}
 \noindent
The compressible Navier-Stokes-Poisson (NSP) system is considered in
$\R^3$ in the present paper and the influences of the electric field
of the internal electrostatic potential force governed by the
self-consistent Poisson equation on the qualitative behaviors of
solutions is analyzed. It is observed that the rotating effect of
electric field affects the dispersion of fluids and reduces the time
decay rate of solutions. Indeed, we show that the density of the NSP
system converges to its equilibrium state at the same $L^2$-rate
$(1+t)^{-\frac34}$ or $L^\infty$-rate $(1+t)^{-3/2}$ respectively as
the compressible Navier-Stokes system, but the momentum of the NSP
system decays at the $L^2$-rate $(1+t)^{-\frac14}$ or
$L^\infty$-rate $(1+t)^{-1}$ respectively, which is slower than the
$L^2$-rate $(1+t)^{-\frac34}$ or $L^\infty$-rate $(1+t)^{-3/2}$ for
the compressible Navier-Stokes system
\cite{LW1998,matsumura1980,D2007}. These convergence rates are also
shown to be optimal for the compressible NSP system.
\end{abstract}



\section{Introduction and main results}
\setcounter{equation}{0}
 The compressible Navier-Stokes-Poisson system takes the form of the
Navier-Stokes equations coupled with the self-consistent Poisson
equation and can be used to simulate, for instance in semiconductor
devices, the transport of charged particles under the electric field
of electrostatic potential force 
\cite{MRS1990}. In the present paper we consider the long time
behavior of global strong solutions of the compressible
Navier-Stokes-Poisson system in $\R^3$. To begin with, we study the
initial value problem (IVP) for the compressible
Navier-Stokes-Poisson (NSP) system
 \bgr
 \partial_t\rho +\div{m}=0,         \label{1.1a}\\
 \partial_tm+\div(\mbox{$\frac{m\otimes{m}}{\rho}$})
  +\nabla p(\rho)+ \rho  \Td\Phi
   =\mu \Lp(\mbox{$\frac{m}{\rho}$})
    +(\mu+\nu)\Td(\div(\mbox{$\frac{m}{\rho}$})), \label{1.1b}\\
 -\lambda^2\Lp\Phi =\rho -\bar{\rho},
   \quad \lim_{|x|\to\infty}\Phi(x,t)\to0,   \label{1.1c}\\
 \rho(x,0) =\rho_0(x),\quad
 m(x,0)=m_0(x), \quad x\in\R^3. \label{1.1d}
 \egr
The variables are the density $\rho>0$, the momentum $m$, the
velocity $u=\frac{m}{\rho}$, and the electrostatic potential $\Phi$.
Furthermore, $p=p(\rho)$ is the pressure function. The viscosity
coefficients satisfy  $\mu>0$, $\frac23\mu+\nu\ge0$, and $\lambda>0$
is the Debye length. $\bar\rho>0$ denotes the background doping
profile, and in this paper is  taken as a positive constant for
simplicity.
\par

When there is no external or internal force involved, there are many
results on the the problem of long time behavior of global smooth
solutions to the compressible Navier-Stokes equations. For
multi-dimensional Navier-Stokes equations, the $H^s$ global
existence and time-decay rate of strong solutions are obtained in
whole space first by Matsumura-Nishida
\cite{matsumura1979,matsumura1980} and the optimal $L^p\ (p\geq2)$
decay rate is established by Ponce \cite{P1985}. The long time decay
rate of global solution in multi-dimensional half space or exterior
domain is also investigated for the compressible Navier-Stokes
equations by Kagei-Kobayashi \cite{KK2002,KK2005}, Kobayashi-Shibata
\cite{KS1999}, and Kobayashi \cite{Ko2002}. Therein, the optimal
$L^2$ time-decay rate in
three dimension is established as 
 \be
\|(\rho-\bar\rho,u)(t)\|_{L^2(\R^3)}
 \le
 C(1 +t)^{-\frac34} \label{1.10z}
\ee
with $(\bar\rho,0)$ the constant state, under small initial
perturbation in Sobolev space. These time-decay rates in energy
space reveal the dissipative properties of the solutions, but
provide no information on wave propagation. For this reason, Zeng
\cite{zeng} analyzes the Green functions of one-dimensional
isentropic Navier-Stokes equations and shows the $L^1$-convergence
to the nonlinear Burgers' diffusive wave. To understand the wave
propagation for compressible fluids in multi-dimension, Hoff-Zumbrun
\cite{hoff,H1997} study the Green's function of an artificial
viscosity system associated with the isentropic Navier-Stokes
equations and derived the $L^\infty$ time-decay rate of diffusive
waves. Liu-Wang \cite{LW1998} investigate  carefully the properties
of the Green's function for isentropic Navier-Stokes system and
present an interesting pointwise convergence of solution to
diffusive waves with the optimal time-decay rate in odd dimension
where the important phenomena of the weaker Huygens' principle is
shown due to the stronger dispersion effects in multi-dimensional
odd space. This is generalized later to the full system later in
\cite{D.L.Li2005} where additional new waves are introduced also. To
conclude, the optimal $L^\infty$ time-decay rate  in three dimension
is
 \be
\|(\rho-\bar\rho,u)(t)\|_{L^\infty(\R^3)}
 \le
 C(1 +t)^{-\frac32}. \label{1.10g}
\ee
The long time behavior for general multi-dimensional
hyperbolic-parabolic systems is studied in \cite{kawashima1987} for
multi-dimensional case with the $H^s$ time-decay rate of solutions
obtained, and in \cite{liu taiping} for one-dimensional case with
$L^p$ $(p\geq1)$ time-decay rate.
\par

When additional (exterior or internal) potential force is taken into
granted, the global existence of strong solution and convergence to
steady state are investigated by Matsumura-Nishida
\cite{matsumura1983}.
As well-known, however, the external force does affect the long time
behavior of dynamical solutions. The slower time-decay rate for
isentropic compressible flow is investigated by Deckelnick
\cite{D1992,D1993},
which is improved later by Shibata-Tanaka \cite{ST2003,STpreprint}
and Ukai-Yang-Zhao \cite{Uhuijiang}. The optimal $L^p$ convergence
rate in $\R^3$ is established recently by Duan-Ukai-Yang-Zhao
\cite{D2007} for the non-isentropic compressible flow as
 \bma
&\|(\rho-\tilde{\rho},u,\theta-\theta_\infty)(t)\|_{L^p(\R^3)}\leq
C(1+t)^{-\frac32(1-\frac1p)},\ \ \ 2\leq p\leq 6,
 \ema
where $(\tilde{\rho}, 0,\theta_\infty)$ is related to the
steady-state solution, under the same smallness assumptions on
initial perturbation and the external force. The optimal $L^p-L^q$
convergence rate with $1\le p<6/5$ and $2 \le q\le 6$ in $\R^3$ was
also established by Duan-Liu-Ukai-Yang \cite{D2008} for isentropic
compressible flow with external potential force.
\par

It also should be noted that for the compressible
Navier-Stokes-Poisson system related to the dynamics of
self-gravitating polytropic gas, there are also important progress
recently on the existence of local and global weak solutions
(re-normalized solution), the reader can refer to for instance
\cite{{DFP113-130},{D31-37},{S257-275},{ZZ305-329}}, and references
therein.
\par

However, there are few results to our knowledge on the global
classical solutions of Navier-Stokes-Poisson system especially for
the analysis of large time behavior, besides the local and/or global
existence of re-normalized weak solution in multi-dimension obtained
by Donatelli \cite{DD345-361} and Zhang-Tan \cite{ZZ305-329} for
Cauchy problem with different hypothesis on the pressure-density
function. To this end, we first study the optimal time-decay rate of
the global classical solutions to the Cauchy problem
\ef{1.1a}--\ef{1.1d}.

To be more precisely, the main purpose in this paper is to study the
existence and uniqueness of global classical solutions and in
particular the asymptotic behavior on the Cauchy problem of
Navier-Stokes-Poisson system.
It is observed that the electric field leads to the rotating
phenomena in fluids motion, affects the speed of wave propagation
and reduces the dispersion effect. This makes the Huygens' principle
observed in \cite{LW1998} for compressible Navier-Stokes equations
invalid here for the compressible compressible Navier-Stokes-Poisson
system and causes the slower time-decay rate of the momentum (or
velocity vector field) to the equilibrium state. Indeed, we show in
this paper that the time-decay rate of the density of the
compressible Navier-Stokes-Poisson system converges to the constant
state at the same algebraic time-decay rate in $\R^3$ (namely,
$(1+t)^{-\frac34}$ in $L^2$-norm or $(1+t)^{-3/2}$ in
$L^\infty$-norm respectively) as the compressible Navier-Stokes
system, but the momentum decays at a slower time-rate  in $\R^3$
(namely, $(1+t)^{-\frac14}$ in $L^2$-norm or $(1+t)^{-1}$ in
$L^\infty$-norm respectively) than both the compressible
Navier-Stokes system \cite{matsumura1980,LW1998} and the
compressible Navier-Stokes system with external force
\cite{Uhuijiang,D2007}. These convergence rates are also shown to be
optimal for the compressible NSP system.
To this end, we consider the linearized NSP system for density and
momentum near an equilibrium state and investigate the spectral of
the linear semigroup in terms of the decomposition of wave modes at
lower frequency and higher frequency respectively. This makes it
possible to analyze the influences of the rotating effect of
electric field (caused by the internal electrostatic potential force
of the self-consistent Poisson equation) on the qualitative
behaviors of the global strong solutions, and finally to obtain the
optimal  $L^p$ ($p\in[2,\infty]$) time decay rate of density and
momentum to the original IVP problem~\qef{1.1a}--\qef{1.1d} in terms
of Duhamel's principle.
\par

First, we have the following theorem on the global existence and
large time behavior of classical solution to the IVP
problem~\qef{1.1a}--\qef{1.1d}.
 \opthm{} 
\label{existence}
 Let $p'(\rho)>0$ for $\rho>0$.
Assume that $(\rho_0-\bar{\rho},m_0)\in H^l(\R^3)\cap L^1(\R^3)$,
$l\ge4 $, with $\delta=: \|(\rho_0-\bar{\rho},m_0)\|_{H^l(\R^3)\cap
L^1(\R^3)}$ small. Then, there is a unique global classical solution
$(\rho,m,\Phi)$ of the IVP~\qef{1.1a}--\qef{1.1d} satisfying
 \bgr
 \rho-\bar{\rho}\in C^0(\R_+,H^l(\R^3))\cap C^1(\R_+,H^{l-1}(\R^3)),\\
  m \in C^0(\R_+,H^l(\R^3))\cap C^1(\R_+,H^{l-2}(\R^3)),\\
 \Phi\in C^0(\R_+,L^6(\R^3)),\quad
 \Td\Phi\in C^0(\R_+,H^{l+1}(\R^3)),
 \egr
and
 \bgr
 \|\partial_x^k(\rho-\bar{\rho})(t)\|_{L^2(\R^3)}
 \leq
  C(1+t)^{-\frac34-\frac k2}
  \|(\rho_0-\bar{\rho},m_0)\|_{H^l(\R^3)\cap L^1(\R^3)},  \label{L2-a}
  \\
 \|\partial_x^km(t)\|_{L^2(\R^3)}
  \leq
  C(1+t)^{-\frac14-\frac k2}
  \|(\rho_0-\bar{\rho},m_0)\|_{H^l(\R^3)\cap L^1(\R^3)},
 \label{L2-c}
\\
 \|\partial_x^{k}\nabla\Phi(t)\|_{L^2(\R^3)}
 \leq
 C(1+t)^{-\frac14-\frac k2}
 \|(\rho_0-\bar{\rho},m_0)\|_{H^l(\R^3)\cap L^1(\R^3)},  \label{L2-b}
 \egr
for $k=0,1$,  where $C>0$ is a positive constant independent of
time.
 \clthm

It should be noted that the time decay rates above are optimal.
Indeed, we shall establish the lower bound time decay rate for the
global solution.

\opthm{}
 \label{Lower-bound}
Let $p'(\rho)>0$ for $\rho>0$. Assume $(\rho_0-\bar{\rho},m_0)\in
H^l(\R^3)\cap L^1(\R^3)$, $l\geq4,$ with $\delta:=
\|(\rho_0-\bar{\rho},m_0)\|_{H^l(\R^3)\cap L^1(\R^3)}$ small enough.
Denote $n_0=:\rho_0-\bar{\rho}$ and assume that the Fourier
transform $\widehat{n}_0=\mathcal{F}({n_0})$ satisfies
$|\widehat{n}_0(\xi)|>c_0>0$ for $0\leq|\xi|\ll1$ with $c_0>0$ a
constant.
Then, the global solution $(\rho,m,\Phi)$ given by
Theorem~\ref{existence} satisfies for $t\ge t_0$ with  $t_0>0$ a
sufficiently large time that
 \bgr
 c_{1}(1+t)^{-\frac34}\leq
 \|(\rho-\bar{\rho})(t)\|_{L^2(\R^3)}
 \leq C(1+t)^{-\frac34},\label{xinde73}
\\
 c_{1}(1+t)^{-\frac14}
 \leq
 \|m(t)\|_{L^2(\R^3)}
 \leq
 C(1+t)^{-\frac14},\label{xinde7}
\\
  c_{1}(1+t)^{-\frac14}\leq
  \|\nabla\Phi(t)\|_{L^2(\R^3)}
  \leq C(1+t)^{-\frac14},\label{xinde72}
 \egr
where $c_{1}, C>0$ are positive constants independent of time.
\clthm

With additional regularity given for the initial data, we can also
prove the optimal $L^p$ time decay rate for the global classical
solution.

\opthm{}
 \label{point-rate}
Let $p'(\rho)>0$ for $\rho>0$. Assume  $(\rho_0-\bar{\rho},m_0)\in
H^l(\R^3)\cap L^1(\R^3)$, $l\geq5$, with $\delta=:
\|(\rho_0-\bar{\rho},m_0)\|_{H^l(\R^3)\cap L^1(\R^3)}$ small enough.
Then, there is a unique global classical solution $(\rho,m,\Phi)$ of
the IVP~\qef{1.1a}--\qef{1.1d} satisfying
 \bgr
 \|(\rho-\bar{\rho})(t)\|_{L^p(\R^3)}
 \leq
 C(1+t)^{-\frac32(1-\frac1p)}
 \|(\rho_0-\bar{\rho},m_0)\|_{H^l(\R^3)\cap L^1(\R^3)},\label{zhudian1}
\\
 \|m(t)\|_{L^p(\R^3)}
\leq
 C(1+t)^{-\frac32(1-\frac1p)+\frac12}
 \|(\rho_0-\bar{\rho},m_0)\|_{H^l(\R^3)\cap L^1(\R^3)},\label{zhudian2}
\\
 \|\nabla\Phi(t)\|_{L^p(\R^3)}
 \leq
 C(1+t)^{-\frac32(1-\frac1p)+\frac12}
 \|(\rho_0-\bar{\rho},m_0)\|_{H^l(\R^3)\cap L^1(\R^3)},  \label{zhudian}
 \egr
for $p\in[2,\infty]$.
\clthm

\begin{remark}
It should be noted that given the same kind of initial data, the
strong solution of the compressible Navier-Stokes equations exists
globally in time and the momentum  decays at the algebraic rate
$(1+t)^{-\frac34}$ in $L^2$-norm or $(1+t)^{-\frac32}$ in
$L^\infty$-norm
respectively~\cite{{matsumura1979},{hoff},{H1997},{LW1998},{D.L.Li2005}},
while the time-decay rate \eqref{L2-a}--\eqref{L2-b} in
Theorem~\ref{existence} and \eqref{zhudian1}--\eqref{zhudian} in
Theorem~\ref{point-rate} implies that the momentum of the
compressible Navier-Stokes-Poisson system decays at the slower
time-decay rate $(1+t)^{-\frac14}$ in $L^2$-norm or  $(1+t)^{-1}$ in
$L^\infty$-norm. This is caused by the coupling of the electric
field and velocity vector field through the Poisson equation, which
also destroys the usual acoustic wave propagation for classical
compressible viscous flow. As one can see in
Theorem~\ref{Lower-bound} that the time decay rate
$(1+t)^{-\frac14}$ of the momentum in $L^2$-norm is optimal.
\end{remark}

\begin{remark} It is also natural to show that the
$L^\infty$-time decay rate established in Theorem~\ref{point-rate}
is optimal. To this end, it is possible to show that the difference
$(\rho-\bar\rho,m,\nabla\Phi)-e^{tB}U_0$ decays more faster than the
corresponding rates in \ef{L2-a}--\ef{zhudian}. In general, however,
it can be carried by the analysis on the fundamental solutions
(Green's function) of the linearized Navier-Stokes-Poisson system in
the framework of Liu-Wang~\cite{LW1998} under modification.
\end{remark}
\begin{remark}
We also note that, the similar properties also hold for the
non-isentropic NSP system, which is left in the incoming paper.
\end{remark}

\bigskip
\noindent{\bf Notations}: In the following part of the paper, $C>0$
and $c_i>0$ with $i\ge 1$ an integer denote the generic positive
constant independent of time.

\bigskip

The paper is arranged as follows. In
section~\ref{spectral-analysis}, we apply the spectral analysis to
the semigroup for the linearized NSP system. We establish the $L^2$
time decay rate of the global solutions for both linearized and
nonlinear NSP system in section~\ref{L2-theory}. In
section~\ref{Lp-theory}, we show the $L^p$ time decay rate.

\section{Spectral analysis of semigroup}
\setcounter{equation}{0}
 \label{spectral-analysis}
Let us consider the IVP problem for the linearized
Navier-Stokes-Poisson system
\bgr
 \dt{n}+\div{m}=0, \label{lin1a}\\
 \dt{m}+ c^2\nabla{n}
   + \lambda^{-2}\Td(-\Lp)^{-1}{n}
   -\mu \Lp{m} -(\mu+\nu)\Td(\div{m})=0, \label{lin1b}\\
 \Phi = \lambda^{-2}(-\Lp)^{-1}n,
   \quad \lim_{|x|\to\infty}\Phi(x,t)\to0, \label{lin1c}\\
 n(x,0) =n_0(x)=:\rho_0(x),\quad
 m(x,0)=m_0(x), \quad x\in\R^3. \label{lin1d}
\egr
In terms of the semigroup theory for evolutional equation, the
solutions $(\bar{n},\bar{m})$ of linear IVP
problem~\ef{lin1a}--\ef{lin1d} can be expressed for
$\bar{U}=(\bar{n},\bar{m})^t$ as
 \be
\bar{U}_t=B\bar{U},\quad \bar{U}(0)=U_0,\quad t\ge 0,  \label{lin2}
 \ee
which gives rise to
 \be
  \bar{U}(t)=S(t)U_0=:e^{tB}U_0,\quad t\ge 0.
\label{lin2b}
 \ee

%
What left is to analyze the differential operator $B$ in terms of
its Fourier expression $A$ and show the long time properties of the
semigroup $S(t)$. Applying the Fourier transform to system
\ef{lin2}, we have
 \be
  \dt{\widehat{U}}= A(\xi) \widehat{U},
\quad
 \widehat{U}(0)=\widehat{U}_0,  \label{lin3}
 \ee
where $\widehat{U}(t)=\widehat{U}(\xi,t)=\mathcal{F}U(\xi,t)$,
$\xi=(\xi_1,\xi_2,\xi_3)^t$ and $A(\xi)$ is defined as
 \be
A(\xi)=\left(\begin{matrix}
      0
 &  -i\xi^t
 \\
   -i\xi(c^2 +  \lambda^{-2}|\xi|^{-2})
 & -\mu |\xi|^{2}I_{3\times 3} -(\mu+\nu)\xi\otimes\xi
 \end{matrix}\right).                  \label{op2}
 \ee
The eigenvalues of the matrix $A$ are computed  from the determinant
 \be
 det (A(\xi)-\lambda I)
 =(\lambda+(2\mu +\nu)|\xi|^2)^2
  (\lambda^2+(2\mu +\nu)|\xi|^2\lambda
  +(c^2+ \lambda^{-2}|\xi|^{-2})|\xi|^2)=0,
\label{determinant}
 \ee
which implies
 \bma
 \lambda_0=&-\mu|\xi|^2,\quad (double)\label{lam0}
\\[2mm]
 \lambda_+
        =&-(\mu+\mbox{$\frac12$}\nu)|\xi|^2
            +\mbox{$\frac12$}i\sqrt{
             4(c^2|\xi|^{2}+ \lambda^{-2})
            -(2\mu +\nu)^2|\xi|^4},\label{lam+}
\\[2mm]
 \lambda_-=&-(\mu+\mbox{$\frac12$}\nu)|\xi|^2
            -\mbox{$\frac12$}i\sqrt{
             4(c^2|\xi|^{2}+ \lambda^{-2})
            -(2\mu +\nu)^2|\xi|^4}.\label{lam-}
 \ema
The semigroup $e^{tA}$ is expressed as
 \be
e^{tA}=e^{\lambda_+t}P_+ + e^{\lambda_-t}P_- + e^{\lambda_0t}P_0
 \ee
where the project operators $P_0,P_\pm$ can be computed as
 \be
 P_i=\mbox{$ \prod_{j\neq i}\frac{A(\xi)-\lambda_jI}{\lambda_i-\lambda_j}$}.
  \label{proj1}
 \ee
By a direct computation, we can verify the exact expression  the
Fourier transform $\widehat{G}(\xi,t)$ of Green's function
$G(x,t)=e^{tB}$  as
\bma
\widehat{G}(\xi,t)=:\,&\, e^{tA}=e^{\lambda_+t}P_+ + e^{\lambda_-t}P_- +
e^{\lambda_0t}P_0\\[2mm]
=& \left(\begin{matrix}
    \frac{\lambda_+e^{\lambda_-t}-\lambda_-e^{\lambda_+t}}
         {\lambda_+-\lambda_-}
 &  -\frac{ i\xi^t(e^{\lambda_+t}- e^{\lambda_-t})}
         {\lambda_+-\lambda_-}
 \\[4mm]
   -\frac{ i\xi c^2(e^{\lambda_+t}- e^{\lambda_-t})}
         {\lambda_+-\lambda_-}
 &  e^{-\lambda_0t}(I- \frac{\xi\otimes\xi}{|\xi|^2})
    +\frac{\xi\otimes\xi}{|\xi|^2}
    \frac{(\lambda_+e^{\lambda_+t}- \lambda_-e^{\lambda_-t})}
         {\lambda_+-\lambda_-}
 \end{matrix}\right) \nnm\\[2mm]
 &+\mbox{$ \frac{e^{\lambda_+t}- e^{\lambda_-t}}
                {\lambda_+-\lambda_-}
 \left(\begin{matrix}
    0
 &  0
 \\
   - \frac{i\xi}{\lambda^{2}|\xi|^{2}}
      & 0
 \end{matrix}\right)$}. \label{green_function}
\ema

To derive the long-time decay rate of solutions whatever in $L^2$
framework or on point-wise estimate, we need to verify the
approximation of the Green's function $G(x,t)$ for both lower
frequency and high frequency. In terms of the definition of the
eigenvalues \ef{lam0}--\ef{lam-}, we are able to obtain that it
holds for $|\xi|\ll1$ that
 \bgr
 \mbox{$\frac{\lambda_+e^{\lambda_-t}-\lambda_-e^{\lambda_+t}}
             {\lambda_+-\lambda_-}
 \sim
   e^{-(\mu+\frac12\nu)|\xi|^2t} \left[\cos(bt)
  + (\mu+\mbox{$\frac12$}\nu)\frac{\sin(bt)}{b}|\xi|^2\right]$},\quad
  |\xi|\ll1,\label{xs1}
\\
  \mbox{$  \frac{\lambda_+e^{\lambda_+t}-\lambda_-e^{\lambda_-t}}
         {\lambda_+-\lambda_-}
 \sim
   e^{-(\mu+\frac12\nu)|\xi|^2t}\left[ \cos(bt)
  - (\mu+\mbox{$\frac12$}\nu)\frac{\sin(bt)}{b}|\xi|^2\right]$},\quad
  |\xi|\ll1,\label{xs2}
\\
 \mbox{$ \frac{e^{\lambda_+t}- e^{\lambda_-t}}
         {\lambda_+-\lambda_-}
 \sim \frac{\sin(tb)}{b}e^{-(\mu+\frac12\nu)t|\xi|^2}$},
 \quad
 |\xi|\ll1,\label{xs3}
\egr
where
 \be
 b= \mbox{$\frac12$}\sqrt{4(c^2|\xi|^{2}+\lambda^{-2})-(2\mu +\nu)^2|\xi|^4}
  \sim   (\lambda^{-1}+\mbox{$\frac{\lambda c^2}{2}$}|\xi|^{2})
   +\mathcal{O}(|\xi|^{4}), \quad |z|\ll1
 \ee

\begin{remark}
For the compressible Navier-Stokes equations, we have
$b=c|\xi|+\mathcal{O}(|\xi|^2)$ for $|\xi|\ll1$. Thus, the dominant
behavior of wave propagation of Navier-Stokes-Poisson equations at
lower frequency is different from the compressible Navier-Stokes.
The so-called Huygens' principle is invalid here.
\end{remark}
\par
\smallskip

To enclose the estimates, we also need to deal with the high
frequency $|\xi|\gg1$. In terms of the definition of the eigenvalues
\ef{lam0}--\ef{lam-}, we are able to analyze the eigenvalues for
$|\xi|\gg1$.
Indeed, we have the leading orders of the eigenvalues for
$|\xi|\gg1$ as
 \bma
 \lambda_0\sim &-\mu|\xi|^2,\label{lamda0}
\\
 \lambda_+\sim&-(2\mu +\nu)|\xi|^2
            +\frac{c^2}{2\mu +\nu}+\mathcal{O}(|\xi|^{-1}),\label{lamda+}
\\
 \lambda_-\sim&-\frac{c^2}{2\mu +\nu}+\mathcal{O}(|\xi|^{-1}).\label{lamda-}
\ema
This approximation gives the leading order terms of the elements of
Green's function as follows
 \bgr\bln
 \mbox{$  \frac{\lambda_+e^{\lambda_-t}-\lambda_-e^{\lambda_+t}}
         {\lambda_+-\lambda_-}$}
 =\,&
  \mbox{$\frac12e^{\lambda_+t}\left[1+e^{(\lambda_--\lambda_+)t}\right]
  + \frac{a}{2b}e^{\lambda_+t}\left[1-e^{(\lambda_--\lambda_+)t}\right]$}
 \\
 \sim\
  & \mathcal{O}(1)e^{-R_0t},\ R_0>0,\quad
  |\xi|\geq \eta,
\eln
\\
  \mbox{$   \frac{\lambda_+e^{\lambda_+t}-\lambda_-e^{\lambda_-t}}
         {\lambda_+-\lambda_-}
 = \,
   \frac{\lambda_+}{2b}e^{\lambda_+t}
    \left[1-e^{(\lambda_--\lambda_+)t}\right]$}
 \sim\,
    \mathcal{O}(1)e^{-R_0t},\ R_0>0,\quad
  |\xi|\geq \eta,
\\
 \mbox{$  \frac{e^{\lambda_+t}- e^{\lambda_-t}}
         {\lambda_+-\lambda_-}
 =
  \frac{1}{2b}e^{\lambda_+t}
  \left[1-e^{(\lambda_--\lambda_+)t}\right]$}
 \sim\,
    \mathcal{O}(1)e^{-R_0t},\ R_0>0,\quad
  |\xi|\geq \eta,
 \egr
where $R_0$ and $\eta$ are some positive constants.

\begin{remark}  The dominant behavior of eigenvalues at higher
frequency for the compressible Navier-Stokes-Poisson system is the
same as the compressible Navier-Stokes.
\end{remark}

\section{$L^2$-time decay rate }
\label{L2-theory} \setcounter{equation}{0}
\subsection{$L^2$ decay rate for linear semigroup}
With the help of the formula \ef{green_function} for Green's
function in Fourier space and the asymptotical analysis on its
elements, we are able to establish the $L^2$ time decay rate.
Indeed, we have the $L^2$-time decay rate of the global strong
solution to the IVP problem for the linearized Navier-Stokes-Poisson
system \ef{lin1a}--\ef{lin1d} as follows.

\opprop{\rm \textbf{\boldmath ($L^2$-theory)}}
 \label{linear_theory_L2-a}
%
Let $U_0=(n_0,m_0)\in H^l(\R^3)\cap L^1(\R^3)$, $l\ge 4$, and denote
$(\bar{n}(t),\bar{m}(t))=:\bar{U}(t)$. Then,
$(\bar{n},\bar{m},\bar{E})$ with $\bar{E}=\Td(-\Lp)^{-1}\bar{n}$
solves the IVP~\eqref{lin1a}--\eqref{lin1d} and satisfies  for
$0\leq k\leq l$ that
 \be \left.\bln
   \|\dx^k \bar{n}(t)\|_{L^2(\R^3)}
 &\le
  C(1+t)^{-\frac34-\frac{k}2}(\|U_0\|_{L^1(\R^3)}
  +\|\partial_x^kU_0\|_{ L^2(\R^3)}),
  \\
 \|\dx^k(\bar{E},\bar{m})(t)\|_{L^2(\R^3)}
 &\le
  C(1+t)^{-\frac14-\frac{k}2}(\|U_0\|_{L^1(\R^3)}
  +\|\partial_x^kU_0\|_{L^2(\R^3)}).
\eln \ \ \right\}                                \label{linear_L2}
\ee
\clprop

\textbf{Proof:} We are going to verify that in $L^2$ norm, it holds
\be
  \|U(t)\|^2_{L^2(\R^3_x)}
 =\|\widehat{G}(\cdot,t)\widehat{U}_0\|^2_{L^2(\R^3_\xi)}
 =\|\widehat{\bar{n}}(t)\|^2_{L^2(\R^3)}+\|\widehat{\bar{m}}(t)\|^2_{L^2(\R^3)}.
\ee
and
\be\bln
  \|\dx^kU(t)\|^2_{L^2(\R^3_x)}
 =&\||\cdot|^k\widehat{U}(\cdot,t)\|^2_{L^2(\R^3_\xi)}
 =\||\cdot|^k\widehat{G}(\cdot,t)\widehat{U}_0\|^2_{L^2(\R^3_\xi)}\\
 =&\||\cdot|^k\widehat{\bar{n}}(\xi,t)\|^2_{L^2(\R^3)}
   +\||\cdot|^k\widehat{\bar{m}}(\xi,t)\|^2_{L^2(\R^3)}.
\eln\ee
An easy computation together with the formula  \ef{green_function}
of the Green's function $\widehat{G}(\xi,t)$ gives
\bma
 \widehat{\bar{n}}(\xi,t)
 =& \mbox{$\frac{\widehat{n}_0(\lambda_+e^{\lambda_-t}-\lambda_-e^{\lambda_+t})}
         {\lambda_+-\lambda_-}
     -\frac{i\xi\cdot\widehat{m}_0(e^{\lambda_+t}- e^{\lambda_-t})}
         {\lambda_+-\lambda_-}$}
 \nnm\\
\sim
 & \left\{\begin{aligned}
  & e^{-(\mu+\frac12\nu)|\xi|^2t}
    \mbox{$
      \left(
        \cos(bt)
       +(\mu+\mbox{$\frac12$}\nu)\frac{\sin(bt)}{b}|\xi|^2\right)\widehat{n}_0
       $}
   \\
  &\quad
   - \mbox{$i\xi\cdot\widehat{m}_0 \frac{\sin(bt)}{b}
     e^{-(\mu+\frac12\nu)|\xi|^2t}$}, \quad |\xi|\ll1,\\
 &  \mathcal{O}(1)e^{-R_0t}(|\widehat{n}_0|+|\widehat{m}_0|),\quad |\xi|\gg1,
\end{aligned}\right.  \nnm
\\
\sim
 & \left\{\begin{aligned}
  &\mathcal{O}(1)e^{-(\mu+\frac12\nu)|\xi|^2t}(|\widehat{n}_0|+|\widehat{m}_0|),
   \quad |\xi|\ll1,\\
 &  \mathcal{O}(1)e^{-R_0t}(|\widehat{n}_0|+|\widehat{m}_0|),\quad |\xi|\gg1,
\end{aligned}
\right.       \label{for-n}
 \ema
with $R_0>0$ a constant here and below, and
 \bma
 \widehat{\bar{m}}(\xi,t)
 =& -\mbox{$\frac{i\xi \widehat{n}_0c^2(e^{\lambda_+t}- e^{\lambda_-t})}
         {\lambda_+-\lambda_-}$}
   +\left[
       \mbox{$\frac{(\lambda_+e^{\lambda_+t}- \lambda_-e^{\lambda_-t})}
            {\lambda_+-\lambda_-}$}
       -e^{-\lambda_0t}
    \right]
     \mbox{$\frac{\xi(\xi\cdot\widehat{m}_0)}{|\xi|^2}$}\nnm\\
  & +\,\widehat{m}_0e^{-\lambda_0t}
   -\mbox{$\frac{i\xi\widehat{n}_0}{ \lambda^{2}|\xi|^2}
           \frac{(e^{\lambda_+t}- e^{\lambda_-t}) }
         {\lambda_+-\lambda_-}$}  \nnm
 \\
\sim
 & \left\{\begin{aligned}
  &+e^{- \mu|\xi|^2t}\left[ e^{-\frac12\nu|\xi|^2t}
      \left(
        \cos(bt)
       -(\mu+\mbox{$\frac12$}\nu)\mbox{$\frac{\sin(bt)}{b}|\xi|^2$}\right)
      -1\right]\mbox{$\frac{\xi(\xi\cdot\widehat{m}_0)}{|\xi|^2}$}
   \\
  & \quad -i\xi\widehat{n}_0(c^2
    +\mbox{$\frac{1}{\lambda^{2}|\xi|^2})\frac{\sin(bt)}{b}$}
      e^{-(\mu+\frac12\nu)|\xi|^2t}
    +\,\widehat{m}_0e^{- \mu|\xi|^2t},
  \  |\xi|\ll1,
 \\
 &  \mathcal{O}(1)e^{-R_0t}(|\widehat{n}_0|+|\widehat{m}_0|),\quad |\xi|\gg1,
\end{aligned}
\right.
\nnm \\
\sim
 & \left\{\begin{aligned}
  &\mathcal{O}(1)e^{-\mu|\xi|^2t}(|\widehat{n}_0|+|\widehat{m}_0|)
        +\mathcal{O}(1)\mbox{$\frac{|\widehat{n}_0|}{|\xi|}e^{-\mu|\xi|^2t}$},
  \quad |\xi|\ll1,
  \\
 &  \mathcal{O}(1)e^{-R_0t}(|\widehat{n}_0|+|\widehat{m}_0|),\quad
 |\xi|\gg1,
\end{aligned}
\right.             \label{xinde9}
 \ema
where and below $\eta>0$ denotes a small but fixed constant.
Therefore, we have the $L^2$-decay rate for $(\bar n,\bar m)$ as
\be
\bln
 \|\widehat{\bar{n}}(t)\|^2_{L^2(\R^3)}
 =&
  \int_{|\xi|\le \eta}|\widehat{\bar{n}}(\xi,t)|^2d\xi
   +\int_{|\xi|\ge \eta} |\widehat{\bar{n}}(\xi,t)|^2d\xi\\
\le
  &  C\int_{|\xi|\le \eta}
    e^{-(2\mu+\nu)|\xi|^2t}(|\widehat{n}_0(\xi)|^2+|\widehat{m}_0(\xi)|^2)d\xi\\
   &+  C\int_{|\xi|\ge \eta}
    e^{-R_0t}(|\widehat{n}_0(\xi)|^2+|\widehat{m}_0(\xi)|^2)d\xi\\
\le
  &C(1+t)^{-\frac32}\|(n_0,m_0)\|^2_{L^2(\R^3)\cap L^1(\R^3)},
\eln
\ee
and
\be
\bln
 \|\widehat{\bar{m}}(t)\|^2_{L^2(\R^3)}
 =&
  \int_{|\xi|\le \eta}|\widehat{\bar{m}}(\xi,t)|^2d\xi
   +\int_{|\xi|\ge \eta} |\widehat{\bar{m}}(\xi,t)|^2d\xi
 \\
\le
  & C \int_{|\xi|\le \eta}
    e^{-2\mu|\xi|^2t}
    (|\widehat{n}_0(\xi)|^2(1+|\xi|^{-2})+|\widehat{m}_0(\xi)|^2)d\xi
  \\
   &+ C \int_{|\xi|\ge \eta}
    e^{-R_0t}(|\widehat{n}_0(\xi)|^2+|\widehat{m}_0(\xi)|^2)d\xi
  \\
\le
  &C(1+t)^{-\frac12}\|(n_0,m_0)\|^2_{L^2(\R^3)\cap L^1(\R^3)},
\eln
 \ee
and the $L^2$-decay rate on the derivatives of $(\bar n,\bar m)$ as
\be
\bln
 \|\widehat{\dx^k\bar{n}}(t)\|^2_{L^2(\R^3)}
 =&
  \int_{|\xi|\le \eta}|\xi|^{2k}|\widehat{\bar{n}}(\xi,t)|^2d\xi
   +\int_{|\xi|\ge \eta} |\xi|^{2k}|\widehat{\bar{n}}(\xi,t)|^2d\xi\\
\le
  & C \int_{|\xi|\le \eta}
    e^{-(2\mu+\nu)|\xi|^2t}
     |\xi|^{2k}(|\widehat{n}_0(\xi)|^2+|\widehat{m}_0(\xi)|^2)d\xi\\
   &+ C \int_{|\xi|\ge \eta} e^{-R_0t}|\xi|^{2k}
    (|\widehat{n}_0(\xi)|^2+|\widehat{m}_0(\xi)|^2)d\xi\\
\le
  &C(1+t)^{-\frac32-k}
   ( \|(n_0,m_0)\|^2_{L^1(\R^3)}
    +\|(n_0,m_0)\|^2_{H^k(\R^3)}),
\eln
\ee
and
\be
\bln
 \|\widehat{\dx^k\bar{m}}(t)\|^2_{L^2(\R^3)}
 =&
  \int_{|\xi|\le \eta}|\widehat{\bar{m}}(\xi,t)|^2d\xi
   +\int_{|\xi|\ge \eta} |\widehat{\bar{m}}(\xi,t)|^2d\xi\\
\le
  & C \int_{|\xi|\le \eta}
    e^{-2\mu|\xi|^2t}|\xi|^{2k}
     (|\widehat{n}_0(\xi)|^2(1+|\xi|^{-2})+|\widehat{m}_0(\xi)|^2)d\xi\\
   &+ C \int_{|\xi|\ge \eta}
    e^{-R_0t}|\xi|^{2k}(|\widehat{n}_0(\xi)|^2+|\widehat{m}_0(\xi)|^2)d\xi\\
\le
  &C(1+t)^{-\frac12-k}( \|(n_0,m_0)\|^2_{L^1(\R^3)}
    +\|(n_0,m_0)\|^2_{H^k(\R^3)}),
\eln
 \ee
for $1\le k\le l$. The estimates on the $\bar{E}$ is obtained via
the expression \ef{lin1c} for $ 0\le k\le l$ as
 \be
 \bln
\|\dx^k\bar{E}\|^2_{L^2(\R^3)}
 =&C\|\dx^k\Td(-\Lp)^{-1}\bar{n}\|^2_{L^2(\R^3)}
 =
 C\||\cdot|^{k-1}\widehat{\bar{n}}(t)\|^2_{L^2(\R^3)}\\
\le
  &C(1+t)^{-\frac12-k}( \|(n_0,m_0)\|^2_{L^1(\R^3)}
    +\|(n_0,m_0)\|^2_{H^k(\R^3)}).
\eln
 \ee
The proof of the Proposition~\ref{linear_theory_L2-a} is completed.
\hfill$\square$

\bigskip

It should be noted that the $L^2$-decay rates derived above are
optimal. Indeed, we have

\opprop{\rm \textbf{}}
 \label{linear_theory_L2-a lower}
Let $U_0=(n_0,m_0)\in H^l(\R^3)\cap L^1(\R^3)$ and assume that the
Fourier transform $\widehat{n}_0=\mathcal{F}({n_0})$ satisfies
$|\widehat{n}_0(\xi)|>c_0>0$ for $0\leq|\xi|\ll1$ with $c_0$ a
constant.
Then,  as $t\rightarrow+\infty$, the solution
$(\bar{n},\bar{m},\bar{E})$ of the IVP~\eqref{lin1a}--\eqref{lin1d}
given by Proposition~\ref{linear_theory_L2-a} satisfies
 \bgr
 c_{1}(1+t)^{-\frac34}\leq
 \|\bar{n}(t)\|_{L^2(\R^3)}
 \leq C(1+t)^{-\frac34},\label{xinde73'}
\\
 c_{1}(1+t)^{-\frac14}
 \leq
 \|\bar{m}(t)\|_{L^2(\R^3)}
 \leq
 C(1+t)^{-\frac14},\label{xinde7'}
\\
  c_{1}(1+t)^{-\frac14}\leq
  \|\bar{E}(t)\|_{L^2(\R^3)}
  \leq C(1+t)^{-\frac14},\label{xinde72'}
 \egr
where $c_{1}, C>0$ are constants which are independent of time.
\clprop

\textbf{Proof:} We only deal with the estimate \ef{xinde7'} for
simplicity, the argument applies to the others. From \ef{xinde9}, we
have
 \bma
 \widehat{\bar{m}}(\xi,t)
 =& -i\xi\widehat{n}_0(c^2+\mbox{$\frac{1}{\lambda^{2}|\xi|^2}$})
      \mbox{$\frac{\sin(bt)}{b}e^{-(\mu+\frac12\nu)|\xi|^2t}$}\nnm\\
  &+e^{- \mu|\xi|^2t}\left[ e^{-\frac12\nu|\xi|^2t}
      \left(
        \cos(bt)
       -(\mu+\mbox{$\frac12$}\nu)
        \mbox{$\frac{\sin(bt)}{b}$}|\xi|^2\right)
      -1\right]
       \mbox{$\frac{\xi(\xi\cdot\widehat{m}_0)}{|\xi|^2}$}\nnm\\
\sim
 & \left\{\begin{aligned}
 & \frac{-i\xi\widehat{n}_0}{\lambda^{2}|\xi|^2}
        \frac{\sin(bt)}{b}
  e^{-(\mu+\frac12\nu)|\xi|^2t}
\\[1mm]
  &\qquad +\mathcal{O}(1)e^{-\mu|\xi|^2t}(|\widehat{n}_0|+|\widehat{m}_0|)
  \equiv T_1+T_2,   \quad |\xi|\ll1,
\\[2mm]
 &  \mathcal{O}(1)e^{-R_0t}(|\widehat{n}_0|+|\widehat{m}_0|)\equiv T,
 \quad |\xi|\gg1,
\end{aligned}
\right.\label{xinde10'}
 \ema
with $R_0>0$ a constant and for lower frequency
$$
 b=\frac12\sqrt{4(c^2|\xi|^{2}+ \lambda^{-2})-(2\mu+\nu)^2|\xi|^4}
  =(\lambda^{-1}+\mbox{$\frac{\lambda c^2}{2}$}|\xi|^{2})
   +\mathcal{O}_1(|\xi|^{4}), \quad |\xi|\ll1.
 $$
in terms of Taylor's expansion. \par

Due to Parseval's  equality $\|\bar{m}\|=\|\widehat{\bar{m}}\|$, it
is enough to estimate the decay rate of$\|\widehat{\bar{m}}(t)\|$.
It is easy to verify
 \bma
 \|\widehat{\bar{m}}(\xi,t)\|^2
= &
 \int_{|\xi|\leq \eta }|\widehat{\bar{m}}(\xi,t)|^2d\xi
   +\int_{|\xi|\geq\eta}|\widehat{\bar{m}}(\xi,t)|^2d\xi
 \nnm\\
\geq &
  \int_{|\xi|\leq \eta}|\widehat{\bar{m}}(\xi,t)|^2d\xi-Ce^{-2R_0 t}
\nnm\\
\geq &
   \int_{|\xi|\leq \eta }\frac12|T_1|^2d\xi
  -\int_{|\xi|\geq \eta}|T_2|^2d\xi-Ce^{-2R_0t}
\nnm\\
\geq&
  \int_{|\xi|\leq \eta}\frac12|T_1|^2d\xi
  -C(1+t)^{-\frac32}-Ce^{-2R_0 t},      \label{xinde11'}
 \ema
where and below $\eta>0$ is a small but fixed constant. By direct
computation we have
 \be
  \int_{|\xi|\leq \eta }|T_1|^2d\xi
 \geq
  C\int_{|\xi|\leq \eta}
   \frac{e^{-(2\mu+\nu)|\xi|^2t}}{|\xi|^2}
   \sin^2(\lambda^{-1}t+\frac{\lambda c^2}{2}|\xi|^2t
          +\mathcal{O}_1(|\xi|^4t))d\xi.       \label{huajian}
  \ee
Applying the mean value formula we have
 \bma
 &\sin(\lambda^{-1} t+\frac{\lambda c^2}{2}|\xi|^2t+\mathcal{O}_1(|\xi|^4t))
\nnm\\
  &=\sin(\lambda^{-1}t+\frac{\lambda c^2}{2}|\xi|^2t)
   +[ \sin(\lambda^{-1} t+\frac{\lambda c^2}{2}|\xi|^2t+\mathcal{O}_1(|\xi|^4t))
     -\sin(\lambda^{-1}t+\frac{\lambda c^2}{2}|\xi|^2t)]
\nnm\\
  &=\sin(\lambda^{-1} t+\frac{\lambda c^2}{2}|\xi|^2t)
    +\mathcal{O}_2(|\xi|^4t),     \nnm \label{hjdxx}
 \ema
and then
 \be
  \sin^2(\lambda^{-1} t+\frac{\lambda c^2}{2}|\xi|^2t+\mathcal{O}_1(|\xi|^4t))
 \geq
   \frac12\sin^2(\lambda^{-1} t+\frac{\lambda c^2}{2}|\xi|^2t)
  -\mathcal{O}_3([|\xi|^4t]^2).         \nnm
 \ee
Substituting above inequality into \ef{huajian} we have
 \bma
\int_{|\xi|\leq \eta}|T_1|^2d\xi
 \geq\,
 & C\int_{|\xi|\leq \eta}
   \frac{e^{-(2\mu+\nu)|\xi|^2t}}{|\xi|^2}
   (\sin^2(\lambda^{-1}t+\frac{\lambda c^2}{2}|\xi|^2t)
    -\mathcal{O}_3([|\xi|^4t]^2))d\xi
\nnm\\
\geq\,
 & C\int_{|\xi|\leq \eta}
    \frac{e^{-(2\mu+\nu)|\xi|^2t}}{|\xi|^2}
     \sin^2(\lambda^{-1}t+\frac{\lambda c^2}{2}|\xi|^2t)d\xi
\nnm\\
 &-C\int_{|\xi|\leq \eta}
     \frac{e^{-(2\mu+\nu)|\xi|^2t}}{|\xi|^2} (|\xi|^4t)^2d\xi \nnm\\
   = &I_1-I_2.\label{huajian'}
  \ema
A direct computation gives rise to
 \bma
 |I_2|
 \le C\int_{|\xi|\leq\eta}
   \frac{e^{-(2\mu+\nu)|\xi|^2t}}{|\xi|^2}(|\xi|^4t)^2d\xi
\leq
  C(1+t)^{-\frac52},            \label{hjd1}
 \ema
and for time $t\ge t_0=:\frac{4R^2}{\eta^2}$ with
$R>\frac{\sqrt{7\pi}}{c\sqrt{\lambda}}$ that
 \bma
 I_1=&\, C\int_{|\xi|\leq \eta}
       \frac{e^{-(2\mu+\nu)|\xi|^2t}}{|\xi|^2}
       \sin^2(\lambda^{-1}t+\frac{\lambda c^2}{2}|\xi|^2t)d\xi
\nnm\\
    =&\,   C t^{-\frac12}\int_{|\zeta|\leq \eta t^{\frac12}}
         e^{-(2\mu+\nu)|\zeta|^2}
        \sin^2(\lambda^{-1}t+\frac{\lambda c^2}{2}|\zeta|^2)d|\zeta|
\nnm\\
    =&\,  Ct^{-\frac12}\left(
        \int_{0}^R +\int_{R}^{\eta t^{\frac12}}
          \right)
        e^{-(2\mu+\nu)r^2}
       \sin^2(\lambda^{-1}t+\frac{\lambda c^2}{2}r^2)dr
\nnm\\
    \geq &\,  c_2(1+t)^{-\frac12}\int_0^R
                e^{-(2\mu+\nu)r^2}
               \sin^2(\lambda^{-1} t +\frac{\lambda c^2}{2}r^2)dr
    \triangleq c_2(1+t)^{-\frac12}F(t)    \label{hjd}
 \ema
with $c_2>0$ a positive constant.
 \par

It is easy to verify that $F(t)$ is a continuous periodic function
of $t$ with the period $\lambda \pi$. It can be also shown that
there is a positive constant $F_{min}>0$ so that
\be
 F(t)\ge F_{min}=:\inf_{\frac{4R^2}{\eta^2}\le s\le t}F(s)>0.
 \label{F_min}
\ee
Indeed, it is trivial to note that for any $t\ge
t_0=:\frac{4R^2}{\eta^2}$ there is an integer $k_0>0$ so that
$t\in[k_0\lambda\pi, (k_0+1)\lambda\pi]$. Thus, to show \ef{F_min},
it is sufficient to deal with $F(t)$ in one time periodic domain,
namely, $t\in[k_0\lambda\pi, (k_0+1)\lambda\pi]$ for some $k_0>0$.
We deal with the case $\lambda^{-1}t\in[k_0\pi,k_0\pi+\frac{\pi}2]$
and $\lambda^{-1}t\in[k_0\pi+\frac{\pi}2,(k_0+1)\pi]$ respectively,
and obtain for $R>\frac{\sqrt{7\pi}}{c\sqrt{\lambda}}$ that
 \bma
 F_{min}=
 &\inf_{ t\in[\lambda k_0\pi+\frac{\pi}2,\lambda(k_0+1)\pi]}
      \int_0^R e^{-(2\mu+\nu)r^2}
       \sin^2(\lambda^{-1} t +\frac{\lambda c^2}{2}r^2)dr
\nnm\\
 \ge & \inf_{ t\in[\lambda k_0\pi+\frac{\pi}2,\lambda(k_0+1)\pi]}
      \left(
       \int_{\frac{\sqrt{2\pi}}{2c\sqrt{\lambda}}}
         ^{\frac{\sqrt{3\pi}}{2c\sqrt{\lambda}}}
      +\int_{\frac{\sqrt{6\pi}}{2c\sqrt{\lambda}}}
         ^{\frac{\sqrt{7\pi}}{2c\sqrt{\lambda}}}
       \right)
       e^{-(2\mu+\nu)r^2}
       \sin^2(\lambda^{-1} t +\frac{\lambda c^2}{2}r^2)dr
\nnm\\
\ge &
  \left\{\begin{aligned}
   & \int_{\frac{\sqrt{2\pi}}{2c\sqrt{\lambda}}}
         ^{\frac{\sqrt{3\pi}}{2c\sqrt{\lambda}}}
       e^{-(2\mu+\nu)r^2}
       \sin^2(\lambda^{-1} t +\frac{\lambda c^2}{2}r^2)dr,
     \quad \lambda^{-1}t\in[k_0\pi,k_0\pi+\frac{\pi}2],
    \\[2mm]
   & \int_{\frac{\sqrt{6\pi}}{2c\sqrt{\lambda}}}
         ^{\frac{\sqrt{7\pi}}{2c\sqrt{\lambda}}}
      e^{-(2\mu+\nu)r^2}
      \sin^2(\lambda^{-1} t +\frac{\lambda c^2}{2}r^2)dr,
     \quad \lambda^{-1}t\in[k_0\pi+\frac{\pi}2,(k_0+1)\pi],
 \end{aligned}\right.
\nnm\\[2mm]
 \ge &
    \mbox{$ \frac{(\sqrt{3\pi}-\sqrt{2\pi})}
                 {2c\sqrt{\lambda}}$}
     \,e^{-\frac{7\pi(2\mu+\nu)}{4c^2\lambda}}
     \sin^2\frac{\pi}{8}>0,\quad
  \lambda^{-1}t\in[k_0\pi+\frac{\pi}2,(k_0+1)\pi]. \label{F_min-2}
 \ema
Thus, it follows from \ef{hjd},  \ef{F_min}   and \ef{F_min-2} that
 \be
 I_1\ge c_1F_{min}(1+t)^{-\frac12}
     =: c_3(1+t)^{-\frac12},   \quad t\ge t_0,       \label{OK1}
 \ee
with $c_3>0$ a constant. \par

Combining \ef{OK1}, \ef{xinde11'}, \ef{huajian'} and \ef{hjd1}, we
obtain the lower bound of time-decay rate for $\bar m$ as
 \be
\|\bar{m}(t)\|_{L^2(\R^3)}=\|\widehat{\bar{m}}(t)\|_{L^2(\R^3)}
 \geq
    c_1(1+t)^{-\frac14},\quad t\ge t_1.  \label{OK1a}
 \ee
for some positive constants $c_1>0$ and $t_1>0$.
\par

The time-decay rate of $\bar n$ and $\bar E$ can be shown in a
similar fashion. Indeed, in view of \ef{for-n} we need only deal
with the lower frequency for the dominating term
$e^{-(\mu+\frac12\nu)|\xi|^2t}\cos (bt)$ to obtain, after a
complicated but straightforward computation, the lower bound of
time-decay rate for $\bar n$ as
 \be
\|\bar{n}(t)\|_{L^2(\R^3)}=\|\widehat{\bar{n}}(t)\|_{L^2(\R^3)}
 \geq
    c_1(1+t)^{-\frac34}.  \label{OK1b}
 \ee
We can re-present the electric field $\bar{E}$ in terms of $\bar{n}$
and the Riesz potential, and deal with the dominating term
$-\frac{i\xi}{|\xi|^2}e^{-(\mu+\frac12\nu)|\xi|^2t}\cos (bt)$ for
lower frequency to obtain the time-decay rate as
 \bma
\|\bar{E}(t)\|_{L^2(\R^3)}=\|\widehat{\bar{E}}(t)\|_{L^2(\R^3)}
 \ge
  c_1(1+t)^{-\frac14},   \label{OK1c}
 \ema
where we recall that
$\widehat{\bar{E}}=-\frac{i\xi}{|\xi|^2}\widehat{\bar{n}}$. The
proof is completed.  \hfill$\square$

\subsection{$L^2$ decay rate for nonlinear system}
\subsubsection{Reformulation of original problem }
Let us reformulate the nonlinear system \ef{1.1a}--\ef{1.1d} for
$(\rho,u)$ near the equilibrium state $(\bar{\rho},0)=(1,0)$. Denote
 \be
  n= \rho-\bar{\rho}, \quad m=m,\quad \Phi=\Phi. \label{pertub}
 \ee
Then, the IVP problem for $(n,m)$ is
 \bgr
 \dt{n} + \div{m}=0, \label{pert1a}
 \\
  \dt{m}  + c^2\nabla{n} +\Td\Phi
    -\mu \Lp{m} -(\mu+\nu)\Td(\div{m}) =-f_0,        \label{pert1b}
\\
  \Phi = \lambda^{-2}(-\Lp)^{-1}n,\quad
   \lim_{|x|\to\infty}\Phi(x,t)\to0, \label{pert1c}\\
 n(x,0) =n_0(x)=:\rho_0(x)-\bar{\rho},\quad
 m(x,0)=m_0(x), \quad x\in\R^3. \label{pert1d}
 \egr
where  $c=c(\bar{\rho})=\sqrt{p'(\bar{\rho})}$ is the sound speed,
and
 \bma
 &f_0=f_0(n,m,\dx{n},\dx{m},\dx^2{n},\dx^2{m})
     =: \div F(n,m,\dx{n},\dx{m}),  \label{inhom}  \\
 &\bln
  F(n,m,\dx{n},\dx{m})=& -\lambda^{-2}\Td(-\Lp)^{-1}{n}\otimes\Td(-\Lp)^{-1}{n}
    +\frac12\lambda^{-2} |\Td(-\Lp)^{-1}{n}|^2 I_{3\times 3}\\
   & +(p(\bar{\rho}+n)-p(\bar{\rho})-c^2{n})I_{3\times 3}\\
   & + \mbox{$\frac{{m}\otimes{m}}{1+n}
    -\mu \Td \(\frac{nm}{1+n}\)
    -(\mu+\nu) \div\(\frac{nm}{1+n}\)I_{3\times 3}$}.\nnm
 \eln
 \ema

Denote
\be
U=(n,m)^t,\quad U_0=(n_0,m_0)^t.\label{vec1}
\ee
We have the equivalent form of system~\ef{pert1a}--\ef{pert1d} in
vector form
\be
\dt{U}=B U+\div{H},\quad U(0)=U_0,
\ee
where the differential operator $B$ is defined as
\be
B=\left(\begin{matrix}
      0
 & - \Td\cdot
 \\
   - c^2\nabla-d^{-2}\nabla(-\Lp)^{-1}
 & -\mu \Lp -(\mu+\nu)\Td\Td\cdot
 \end{matrix}\right)                 \label{op1}
\ee
and the nonlinear term $H$ is expressed  by
\be
H(U,\dx U)=(0, F(U,\dx U))^t.   \label{H1}
\ee
Thus, we can represent the solution in term of the semigroup
\be
U(t)=S(t)U_0 +\int_0^t S(t-s)\div{H}(U,\dx U)ds
\ee
with the semigroup $S(t)$ defined via multiplier through Fourier
transformation
\be
S(t)U=e^{tB}U=\mathcal{F}^{-1}e^{tA(\xi)}\mathcal{F}U,\quad
A(\xi)=\mathcal{F}(B)(\xi),\ \xi\in \R^3.
\ee
\par

To establish the time decay rate of the original nonlinear problem,
we need to decompose the Green's function $G=:e^{tB}$ in terms of
its Fourier transform $\widehat{G}(\xi)$. Indeed, by the formula
\ef{green_function}, we can make the following decomposition for
$(\bar{n},\bar{m})=G* U_0$ as
 \be
 \widehat{\bar{n}}=\widehat{N}\cdot\widehat{
U_0}=(\widehat{\mathcal{N}}+\widehat{\mathfrak{N}})\cdot\widehat{
U_0},\quad \widehat{\bar{m}}=\widehat{M}\cdot\widehat{
U_0}=(\widehat{\mathcal{M}}+\widehat{\mathfrak{M}})\cdot\widehat{
U_0},\label{define}
 \ee
where
 \bma
&\widehat{\mathcal{N}}=\left(\begin{matrix}
    \frac{\lambda_+e^{\lambda_-t}-\lambda_-e^{\lambda_+t}}
         {\lambda_+-\lambda_-}\
 &  0
 \end{matrix}\right)_{1\times4},\ \ \ \widehat{\mathfrak{N}}=\left(\begin{matrix}
   0\
 &  -\frac{ i\xi^t(e^{\lambda_+t}- e^{\lambda_-t})}
         {\lambda_+-\lambda_-}
 \end{matrix}\right)_{1\times4},\label{define1}\\
 &\widehat{\mathcal{M}}=\left(\begin{matrix} -\frac{ i\xi c^2(e^{\lambda_+t}-
e^{\lambda_-t})}
         {\lambda_+-\lambda_-}
 &\ \ \ \ \   e^{-\lambda_0t}(I- \frac{\xi\otimes\xi}{|\xi|^2})
    +\frac{\xi\otimes\xi}{|\xi|^2}
    \frac{(\lambda_+e^{\lambda_+t}- \lambda_-e^{\lambda_-t})}
         {\lambda_+-\lambda_-}
 \end{matrix}\right)_{3\times4},\label{define2}\\
 &\widehat{\mathfrak{M}}=\frac{ \lambda^{-2}(e^{\lambda_+t}- e^{\lambda_-t}) }
         {\lambda_+-\lambda_-}\left(\begin{matrix}
   - \frac{i\xi}{|\xi|^{2}}
      &\ \  0
 \end{matrix}\right)_{3\times4}.\label{define3}
 \ema
And we have the Fourier expression for the electric field $\bar{E}$
as
  \bma
 \widehat{ \bar{E}}&=-\frac{i\xi}{|\xi|^2}\widehat{\bar{n}}
=(-\frac{i\xi}{|\xi|^2}\otimes \widehat{N})\widehat{U}_0
=(-\frac{i\xi}{|\xi|^2}\otimes
\widehat{\mathcal{N}})\widehat{U}_0+(-\frac{i\xi}{|\xi|^2}\otimes
\widehat{\mathfrak{N}})\widehat{U}_0,
  \ema
from which we can define
 \bma
 \widehat{\bar{E}}=\widehat{L}\widehat{U}_0
  =(\widehat{\mathcal{L}}+\widehat{\mathfrak{L}})\widehat{U}_0,\label{define4}
  \ema
where
 \bma
\widehat{\mathcal{L}}=(-\frac{i\xi}{|\xi|^2}\otimes
\widehat{\mathcal{N}}),\ \
\widehat{\mathfrak{L}}=(-\frac{i\xi}{|\xi|^2}\otimes
\widehat{\mathfrak{N}}).
 \ema
It is easy to verify that the global solution $(U,E)$ of the IVP
problem for the  nonlinear Navier-Stokes-Poisson
system~\ef{pert1a}--\ef{pert1d} is
$$
U=(n,m)=S(t)U_0+\int_0^tS(t-\tau)Q(U)d\tau,\quad
E=\Td(-\Lp)^{-1}{n},
$$
with $Q(U)=\nabla\cdot H$, which can also be decomposed as
 \bma
 &n=N\ast U_0+\int_0^t\mathfrak{N}(t-\tau)\ast Q(U)(\tau)d\tau,\label{express n}\\
 &m=M\ast U_0+\int_0^t\mathcal{M}(t-\tau)\ast Q(U)(\tau)d\tau,\label{express m}\\
 &E=L\ast U_0+\int_0^t\mathfrak{L}(t-\tau)\ast Q(U)(\tau)d\tau,\label{express E}
 \ema

\par
By Proposition \ref{linear_theory_L2-a}, we have the following time
decay rate for linear part
\bgr
 \|\partial_x^{\alpha}N\ast
u_0(t)\|_{L^2}\leq C(1+t)^{-\frac34-\frac{|\alpha|}{2}}
(\|u_0\|_{L^1}+\|\partial_x^{\alpha}u_0\|_{L^2}),
 \label{wuqiong1''}
 \\
\|\partial_x^{\alpha}M\ast u_0(t)\|_{L^{2}}\leq
C(1+t)^{-\frac14-\frac{|\alpha|}{2}}
(\|u_0\|_{L^1}+\|\partial_x^{\alpha}u_0\|_{L^2}),
 \label{wuqiong2''}
 \\
\|\partial_x^{\alpha}L\ast u_0(t)\|_{L^{2}}\leq
C(1+t)^{-\frac14-\frac{|\alpha|}{2}}
(\|u_0\|_{L^1}+\|\partial_x^{\alpha}u_0\|_{L^2}),
 \label{wuqiong3''}
 \egr
for $|\alpha|\geqslant0$. Furthermore, in view of \ef{xs1}--\ef{xs3}
and the above definition \ef{define} for $\mathfrak{\widehat{N}},
\mathcal{\hat {M}},\mathfrak{\widehat{L}}$, it is easy to verify for
some constant $c_4>0$ that
 \be |\mathfrak{\widehat{N}}(\xi)|\sim
\mathcal{O}(1)|\xi|e^{-c_4|\xi|^2t},\ \
 |\mathcal{\hat
{M}}(\xi)|\sim \mathcal{O}(1)e^{-c_4|\xi|^2t},\  \
|\mathfrak{\widehat{L}}(\xi)|\sim \mathcal{O}(1)e^{-c_4|\xi|^2t},\ \
\ |\xi|\ll1.
 \ee
Thus, applying the similar argument as in the proof of Proposition
\ref{linear_theory_L2-a}, we are able to obtain after a
straightforward computation (which we omit the details) that
 \bma
&\|\partial_x^{\alpha}\mathfrak{\widehat{N}}\ast u_0(t)\|_{L^2}
 \leq
 C(1+t)^{-\frac32(\frac1q-\frac12)-\frac12-\frac{|\alpha|}{2}}
 (\|u_0\|_{L^q}+\|\partial_x^{\alpha}u_0\|_{L^2}),\ q=1,2,
 \label{wuqiong1'''}
 \\
 & \|\partial_x^{\alpha}\mathcal{\hat{M}}\ast u_0(t)\|_{L^{2}}
 \leq
 C(1+t)^{-\frac32(\frac1q-\frac12)-\frac{|\alpha|}{2}}
 (\|u_0\|_{L^q}+\|\partial_x^{\alpha}u_0\|_{L^2}),\ q=1,2,
 \label{wuqiong2'''}
 \\
 &\|\partial_x^{\alpha}\mathfrak{\widehat{L}}\ast u_0(t)\|_{L^{2}}
 \leq
 C(1+t)^{-\frac32(\frac1q-\frac12)-\frac{|\alpha|}{2}}
( \|u_0\|_{L^q}+\|\partial_x^{\alpha}u_0\|_{L^2}),\ q=1,2.
 \label{wuqiong3'''}
 \ema

\subsubsection{Global existence and $L^2$-decay rate}

We are now ready to prove Theorem~\ref{existence} and
Theorem~\ref{Lower-bound} on the optimal time decay rate of solution
to the IVP problem of the nonlinear Navier-Stokes-Poisson
system~\ef{pert1a}--\ef{pert1d}. From \ef{H1} we have
 \be
 Q(U)=\nabla\cdot H(U,\nabla U)=Q_1+Q_2+Q_3,
 \ee
which implies for smooth solution $(\rho,m)$ satisfying
$\|(\rho-\bar\rho,m)\|_{H^4}<\infty$ that
 \bma
 &Q_1:=Q_1(U)\sim \mathcal{O}(1)(\partial_xm \cdot m), \label{Q1}\\
 &Q_2:=Q_2(U)\sim \mathcal{O}(1)\partial_xH_2,\ \ H_2= \partial_x(m\cdot n),\label{Q2}\\
 &Q_3:=Q_3(U)\sim \mathcal{O}(1)\partial_xn\cdot n
        +\mathcal{O}(1)(n\cdot E).\label{Q3}
 \ema

\underline{\it Framework of short time existence}:  First of all, we
give the local existence theory which can be established in the
framework as \cite{matsumura1980,matsumura1979,matsumura1983}.
Indeed, starting with the equations \ef{1.1a}--\ef{1.1d}, we can
make use of the theorem of contracting map to establish the local
existence. The key point is that the electric field $E$ can be
expressed  by
 \ef{1.1a}, \ef{1.1c} and the Riesz potential as a nonlocal term
$$
 E=E_0+\nabla(-\bigtriangleup^{-1})div\int_0^tmds
  =E_0+\nabla(-\bigtriangleup^{-1})div\int_0^t\rho uds,
$$
which together with the $L^p$ estimates leads to
$$
\parallel\nabla(-\bigtriangleup^{-1})div\int_0^t\rho uds\parallel_{H^k}
\leq C\parallel\int_0^t\rho uds\parallel_{H^k}.
$$
Then, by the standard argument of contracting map theorem as
\cite{matsumura1980,matsumura1979,matsumura1983}, one can obtain the
short time existence of strong solution. The details are omitted.
\par

To extend the short time strong solution to be a global in time
solution, we need to establish the uniform a-priori estimates.
Indeed, under some a-priori assumptions we are able to obtain the
expected estimates with time decay rate for lower order terms, and
finally enclose the  a-priori assumptions. We have

\begin{lemma} \label{xin-lemma}
Under the assumptions of Theorem \ref{existence}, the solution
$(n,m,E)$ with $E=\nabla\Phi$ of the IVP
problem~\eqref{pert1a}--\eqref{pert1d} satisfies for $l=4$ that
 \bman
 &\ \ \|\partial_x^kn(t)\|_{L^2}\leq C(1+t)^{-\frac34-\frac k2},\ \
  k=0,1,\
  \
 & \|\partial_x^2n(t)\|_{L^2}\leq C(1+t)^{-\frac34},\\
 &\ \ \|\partial_x^km(t)\|_{L^2}\leq C(1+t)^{-\frac14-\frac k2},\ \ k=0,1,\
  \
 &\|\partial_x^2m(t)\|_{L^2}\leq C(1+t)^{-\frac34},\\
 & \ \ \|\partial_x^k\Phi(t)\|_{L^2}\leq C(1+t)^{-\frac14-\frac k2},\ \ k=1,2,\
 &\|\partial_x^3\Phi(t)\|_{L^2}\leq C(1+t)^{-\frac34},
 \eman
and for $l=5$ that
 \bman
 &\ \ \|\partial_x^kn(t)\|_{L^2}\leq C(1+t)^{-\frac34-\frac k2},\ \ k=0,1,2,\
  \
 & \|\partial_x^3n(t)\|_{L^2}\leq C(1+t)^{-1},\\
 &\ \ \|\partial_x^km(t)\|_{L^2}\leq C(1+t)^{-\frac14-\frac k2},\ \ k=0,1,2,\
  \
 &\|\partial_x^3m(t)\|_{L^2}\leq C(1+t)^{-1},\\
 & \ \ \|\partial_x^k\Phi(t)\|_{L^2}\leq C(1+t)^{-\frac14-\frac k2},\ \ k=1,2,3,\
 &\|\partial_x^4\Phi(t)\|_{L^2}\leq C(1+t)^{-1},
 \eman
where  $C$ is a positive constant independent of time.
 \end{lemma}
\demo   Suppose that $(n,m,E)\in (H^4)^2\times H^5$ and $(n,m,E)\in
(H^5)^2\times H^6$ with $E=\nabla \Phi$ the electric field
correspond respectively to the strong solutions of the compressible
Navier-Stokes-Poisson system for $t\in[0,T]$ subject to  initial
data in different Sobolev space. Assume that the classical solution
of the IVP~\ef{pert1a}--\ef{pert1d} exists for $t\in[0,T]$ and
denote
 \bma
  \Lambda_{1}(t) :=\sup_{0\leq s\leq t,\,k=0,1}
 \{&\|D_x^kn(s)\|(1+s)^{\frac34+\frac k2}
  +\|D_x^km(s)\|(1+s)^{\frac14+\frac k2}
  +\|D_x^2(n,m)(s)\|(1+s)^{\frac34}
  \nnm\\
 &
  +\|E(s)\|(1+s)^{\frac14}
  +\|(D_x^3n,D_x^4n,D_x^3m,D_x^4m)(s)\|\},
 \label{suppose 1}
 \ema
 and
  \bma\Lambda_{2}(t) :=\sup_{0\leq s\leq t,\, k=0,1,2}
\{&\|D_x^kn(s)\|(1+s)^{\frac34+\frac k2}+\|D_x^3n(s)\|(1+s)
 +\|(D_x^4n,D_x^5n,D_x^4m,D_x^5m)(s)\|\nnm \\
&+\|D_x^3m(s)\|(1+s)^{\frac34}+\|E(s)\|(1+s)^{\frac14}
 +\|D_x^km(s)\|(1+s)^{\frac14+\frac k2}\}.     \label{suppose 2}
 \ema

We claim that it holds for any $t\in[0,T]$ that
 \be
 \Lambda_{1}(t)\leq C\delta,\quad \label{claim 1}
 \Lambda_{2}(t)\leq C\delta, 
 \ee
with $\delta$ defined in Theorem~\ref{existence}. In addition, the
claim  \ef{claim 1} together with the smallness assumption on
$\delta$ are sufficient for us to prove the Lemma~\ref{xin-lemma}
and Theorem~\ref{existence}.

Next, let us prove the claim \ef{claim 1}. It is sufficient to show
the first one in \ef{claim 1} since the arguments can be also
applied to second one in \ef{claim 1} under minor modifications. The
proof of the first one in \ef{claim 1} consists of following three
steps.
\par

\underline{\it Step 1: The basic energy estimates}.\ \ Starting with
\ef{express n}, Proposition~\ref{linear_theory_L2-a},
\ef{wuqiong1''}, \ef{wuqiong1'''}, and the a-priori assumption
\ef{suppose 1}, we have after a complicated but straightforward
computation that
 \bma
  \|(n-N\ast U_0)(t)\|
 &\leq 
 \int_0^t \|\mathfrak{N}(t-\tau)\ast Q(U)(\tau)\|d\tau\nnm\\
 &\leq 
  C\int_0^{ t}(1+t-\tau)^{-\frac34-\frac12}
 (\| Q(U)(\tau)\|+\| Q(U)(\tau)\|_{L^1})d\tau\nnm\\
 &\leq 
 C\int_0^{ t}(1+t-\tau)^{-\frac34-\frac12}
 (\Lambda_{1}(t))^2(1+\tau)^{-1}d\tau\nnm\\
 &\leq 
 C(1+t)^{-1}(\Lambda_{1}(t))^2,\label{n L2 0}
 \ema
where we have made use of \ef{suppose 1} and \ef{Q1}--\ef{Q3} to
estimate the right hand side terms as
 \bma
 (\| Q(U)\|+\| Q(U)\|_{L^1})&\leq \{\| (Q_1+Q_2+Q_3)\|\}
 +\{\|(Q_1+Q_2+Q_3) \|_{L^1}\}\nnm\\
 &\leq C\{\|Dm\|\|m\|_{L^\infty}+\|D^2m\|\|n\|_{L^\infty}
    +\|m\|_{L^\infty}\|D^2n\|\nnm\\
 &\hspace{12mm}+\|n\|\|E\|_{L^\infty}+\|Dn\| \|n\|_{L^\infty}\}\nnm\\
 &\ \ \ +C\{\|Dm\| \|m\|+\|D^2m\| \|n\|+\|Dm\|\|Dn\|+\|m\|
 \|D^2n\|\nnm\\
 &\hspace{12mm}+\|n\|\|E\|+\|Dn\|\|n\|\}\nnm\\
 &\leq C (1+t)^{-\frac32}(\Lambda_{1}(t))^2
    +C(1+t)^{-1}(\Lambda_{1}(t))^2.\label{Q L2 1}
 \ema
In a similar fashion, we are able to estimate the high order
derivatives for density as follows. Indeed, in terms of
\ef{wuqiong1'''}, the H\"older's inequality  and Nirenberg's
inequality
$$
 \|u\|_{L^\infty}\leq
C\|Du\|^{\frac12}\|D^2u\|^{\frac12}
$$
and the fact
 \be
 \|D^kE\|\leq C\|D^{k-1}n\|,\ \ k\geq1\label{E L2 estimate}
 \ee
due to the Riesz potential representation, we can estimate the term
$Dn$ as
 \bma
 \|Dn(t)\|&\leq \|D(N\ast U_0)(t)\|+\int_0^t
  \|D(\mathfrak{N}(t-\tau)\ast Q(U))(\tau)\|d\tau
\nnm\\
 &\leq C\delta(1+t)^{-\frac54}
 +\int_0^{\frac t2}\|D(\mathfrak{N}(t-\tau)\ast
 Q(U)(\tau))\|d\tau
 +\int_{\frac t2}^{t}\|D(\mathfrak{N}(t-\tau)\ast Q(U)(\tau))\|d\tau
 \nnm\\
&\leq
  C\delta(1+t)^{-\frac54}
 +C\int_0^{\frac t2}(1+t-\tau)^{-\frac{7}4}
  (\| Q(U)(\tau)\|_{L^1}+\| DQ(U)(\tau)\|)d\tau
 \nnm\\
 &\hspace{5mm}
  +C\int_{\frac t2}^{ t}\|\mathfrak{N}(t-\tau)\ast DQ(U)(\tau)\|d\tau
\nnm\\
 &\leq
   C\delta(1+t)^{-\frac54}
  +C(\Lambda_{1}(t))^2\int_0^{\frac t2}(1+t-\tau)^{-\frac{7}4}
   (1+\tau)^{-1}d\tau
 \nnm\\
 &+C\int_{\frac t2}^{ t}(1+t-\tau)^{-1}
  \|Q(U)(\tau)\|d\tau
\nnm\\
 &\leq
  C\delta(1+t)^{-\frac54}
  +C(\Lambda_{1}(t))^2\int_0^{\frac t2}
   (1+t-\tau)^{-\frac{7}4}(1+\tau)^{-1}d\tau
  \nnm\\
 &+C(\Lambda_{1}(t))^2\int_{\frac t2}^{ t}
(1+t-\tau)^{-1}
 (1+\tau)^{-\frac32}d\tau\nnm\\
 &\leq
   C\delta(1+t)^{-\frac54}
  +C(1+t)^{-\frac54-\varepsilon}(\Lambda_{1}(t))^2,  \label{n L2 1}
 \ema
where and below $\varepsilon>0$ is a small but fixed constant.
As for $D^2n$, it is easy to verify that the nonlinear terms in
right hand side dominating the time-decay rate consists of
$D(D^2(mn))$, $D^2(Dm\cdot m)$ and $D^2(D^2(mn))$, which can be
estimated due to the following facts
 \bgr
 \|D^3n\cdot m(t)\|\leq \|D^3n(t)\|\|m(t)\|_{L^\infty}\leq
 C\delta(1+t)^{-\frac34}\Lambda_{1}(t),\nnm
 \\
 \|D^3m\cdot m(t)\|\leq\|D^3m(t)\|\|m(t)\|_{L^\infty}
 \leq
   C\delta(1+t)^{-\frac34}\Lambda_{1}(t), \nnm
 \\
 \|D^4n\cdot m(t)\|\leq\|D^4n(t)\|\|m(t)\|_{L^\infty}\leq
C\delta(1+t)^{-\frac34}\Lambda_{1}(t).\nnm
 \egr
Thus, we can obtain after a straightforward computation that
 \bma
 \|D^2n(t)\|&\leq \|D^2(N\ast U_0)(t)\|+\int_0^t
 \|D^2(\mathfrak{N}(t-\tau)\ast Q(U))(\tau)\|d\tau\nnm\\
 &\leq C\delta(1+t)^{-\frac74}
 +\int_0^{\frac t2}\|D^2(\mathfrak{N}(t-\tau)\ast
 Q(U)(\tau))\|d\tau\nnm\\
 &\ \ \
 +\int_{\frac t2}^{t}\|D^2(\mathfrak{N}(t-\tau)
 \ast Q(U)(\tau))\|d\tau\nnm\\
&\leq C\delta(1+t)^{-\frac74}+C\int_0^{\frac t2
}(1+t-\tau)^{-\frac94}
 (\| Q(U)(\tau)\|_{L^1}+\|D^2Q(U)(\tau)\|)d\tau\nnm\\
 &\ \ \
 +\int_{\frac t2}^{t}\|(1+t-\tau)^{-\frac32}
 (\|Q(U(\tau))\|+\|D^2Q(U(\tau))\|)d\tau.\nnm\\
 &\leq C\delta(1+t)^{-\frac74}+C(\Lambda_{1}(t))^2\int_0^{\frac t2
}(1+t-\tau)^{-\frac94}
 (1+\tau)^{-\frac34}d\tau\nnm\\
 &+C\delta\Lambda_{1}(t)\int_{\frac t2}^{t}(1+t-\tau)^{-\frac32}
 (1+t)^{-\frac34}d\tau\nnm\\
 &\leq C\delta(1+t)^{-\frac74}+C(\Lambda_{1}(t))^2(1+t)^{-\frac74}
 +C\delta\Lambda_{1}(t)(1+t)^{-\frac34},
 \label{n L2 2}
 \ema
\par

Next, in terms of Proposition~\ref{linear_theory_L2-a}, \ef{express
m}, \ef{wuqiong2''}, \ef{wuqiong2'''}, the a-priori
assumption~\ef{suppose 1}, the H\"older's and Nirenberg's
inequalities, we can prove the time decay rate for $m$ and its
derivatives as follows.
 \bma
 \|(m-M\ast U_0)(t)\|
&\leq 
  \int_0^t \|\mathcal{M}(t-\tau)\ast Q(U)(\tau)\|d\tau\nnm\\
 &\leq 
 C\int_0^{ t}(1+t-\tau)^{-\frac34}
 (\| Q(U)(\tau)\|+\| Q(U)(\tau)\|_{L^1})d\tau\nnm\\
 &\leq 
 C\int_0^{ t}(1+t-\tau)^{-\frac34}
 (\Lambda_{1}(t))^2(1+\tau)^{-1}d\tau\nnm\\
 &\leq 
 C(1+t)^{-\frac14-\varepsilon}(\Lambda_{1}(t))^2,\label{m L2 0}
 \ema
where and below $\varepsilon>0$ is a small but fixed constant, and
 \bma
 \|Dm(t)\|&\leq \|D(M\ast U_0)(t)\|+\int_0^t
 \|D(\mathcal{M}(t-\tau)\ast Q(U)(\tau))\|d\tau\nnm\\
 &\leq C\delta(1+t)^{-\frac34}
 +C\int_0^{ \frac t2}(1+t-\tau)^{-\frac34-\frac12}
 (\| Q(U)(\tau)\|_{L^1}+\| DQ(U)(\tau)\|)(d\tau\nnm\\
 &\ \ \ \
 +C\int_{ \frac t2}^{t}(1+t-\tau)^{-\frac34-\frac12}
 (\| Q(U)(\tau)\|_{L^1}+\| DQ(U)(\tau)\|)d\tau\nnm\\
 &\leq C\delta(1+t)^{-\frac34}
 +C(1+t)^{-\frac34-\varepsilon}(\Lambda_{1}(t))^2,   \label{m L2 1}
 \ema
where we have used \ef{Q L2 1}. As for $D^2m$, we have
 \bma
 \|D^2m(t)\|&\leq \|D^2(M\ast U_0) (t)\|+\int_0^t
 \|D^2(\mathcal{M}(t-\tau)\ast Q(U)(\tau))\|d\tau\nnm\\
 &\leq C\delta(1+t)^{-\frac54}
 +C\int_0^{ \frac t2}(1+t-\tau)^{-\frac{7}4}
 (\| Q(U)(\tau)\|_{L^1}+\| D^2Q(U)(\tau)\|)d\tau\nnm\\
 &\ \ \ \
 +C\int_{ \frac t2}^{t}(1+t-\tau)^{-\frac34-\frac12}
 (\|Q(U)(\tau)\|_{L^1}+\| D^2 Q(U)(\tau)\|)d\tau\nnm\\
 &\leq C\delta(1+t)^{-\frac54}
 +C(1+t)^{-\frac64-\varepsilon}(\Lambda_{1}(t))^2
+C\delta\int_{ \frac t2}^{t}(1+t-\tau)^{-\frac34-\frac12}
 (1+\tau)^{-\frac34}d\tau\nnm\\
 &\leq C\delta\Lambda_{1}(t)(1+t)^{-\frac54}
 +C(1+t)^{-\frac64-\varepsilon}(\Lambda_{1}(t))^2
+C\delta\Lambda_{1}(t)(1+t)^{-\frac34}.\label{m L2 2}
 \ema

What left is to obtain the time-decay rate for $E$ in terms of \ef{E
L2 estimate} and \ef{express E}. Indeed, it is easy to get
 \bma
 \|(E-L\ast U_0)(t)\|
 &\leq 
  \int_0^t\|\mathfrak{L}(t-\tau)\ast Q(U)(\tau)\|d\tau\nnm\\
 &\leq 
 C\int_0^{ t}(1+t-\tau)^{-\frac34}
 (\| Q(U)(\tau)\|+\| Q(U)(\tau)\|_{L^1})d\tau\nnm\\
 &\leq 
 C\int_0^{ t}(1+t-\tau)^{-\frac34}
 (\Lambda_{1}(t))^2(1+\tau)^{-1}d\tau\nnm\\
 &\leq 
 C(1+t)^{-\frac14-\varepsilon}(\Lambda_{1}(t))^2.\label{E L2 0}
 \ema
\par

\underline{\it Step 2: The higher order energy  estimates.} \ To
enclose the a-priori estimates and prove the claim~\ef{claim 1}, we
need to derive the time decay rate of $(n,m)$ with respect to higher
order derivatives as in
\cite{matsumura1980,matsumura1979,matsumura1983}. To this end, we
look on the compressible Navier-Stokes-Poisson
system~\ef{pert1a}--\ef{pert1c} as the compressible Navier-Stokes
with nonlinear inhomogeneous terms related to the electric field
$E$. Indeed, by \ef{1.1a}--\ef{1.1d}, we derive the system for
$(\rho,u)=(1+n,\frac{m}{\rho})$
 \bgr
 \dt{n} + \div{u}=f_1, \label{pert1a'}
 \\
  \dt{u}  +\nabla{n} +E
    -\mu_1 \Lp{u} -\mu_2\Td(\div{u}) =-f_2-f_3,        \label{pert1b'}
\\
  E = \nabla(-\Lp)^{-1}n,\quad \lim_{|x|\to\infty}E\to0, \label{pert1c'}\\
 n(x,0) =n_0(x)=:\rho_0(x)-1,\quad
 u(x,0)=u_0(x)=m_0(x)/\rho_0(x), \quad x\in\R^3,   \label{pert1d'}
  \egr
where
 \be
 \bln
&f_1=f_1(n,u,\dx{n},\dx{u})
  =: -n\div u-\nabla n\cdot u,
 \\
  &f_2=f_2(n,m,\dx{n},\dx{m},\dx^2{n},\dx^2{m})  \\
  &\ \ \  =: (u\cdot\nabla)u+(1-\frac{p'(1+n)}{1+n})\nabla n
  +\mu_1 (\frac{n}{1+n})\Lp{u}+\mu_2(\frac{n}{1+n})\Td(\div{u}), \\
&f_3=-n\nabla\Phi=-nE,
 \eln                   \label{inhom-a}
 \ee
and we have chosen $\bar{\rho}=1,p'(1)=1,\mu=\mu_1,(\mu+\nu)=\mu_2$
and $d=1,$ for simplicity. Notice that, in view of \ef{pert1a'} and
\ef{pert1b'}, there are only two terms different from the classical
Navier-Stokes equations, that is, $E$ and $nE$.
\par

Taking $\int \ef{pert1a'}\times ndx+\int \ef{pert1b'}\cdot udx$ and
integrating the resulted equation by parts,  and making use of the
facts  $E=\nabla\Phi$  and
 \bma
 \int\nabla\Phi\cdot udx
 &=-\int\Phi \nabla\cdot udx
    =\int\Phi n_tdx+\int\Phi f_1dx\nnm\\&
  =-\int\Phi \nabla\cdot E_tdx+\int\Phi f_1dx
 =\int E\cdot E_tdx+\int\Phi f_1dx\nnm\\
 &\geq
     \frac12\frac{d}{dt}\int |E|^2dx
    -\|\Phi \|_{L^6}\| f_1\|_{L^{\frac65}}
  \geq
    \frac12\frac{d}{dt}\int |E|^2dx
   -C\| E \|\| f_1\|_{L^{\frac65}} \label{basic.2}
 \ema
due to the H\"{o}lder's inequality and Sobolev inequality, we can
obtain after a straightforward computation that
 \bma
 &\frac12\frac{d}{dt}\int(|n|^2+|u|^2+|E|^2)(t)dx+C\int|Du(t)|^2dx\nnm\\
 \leq
  & \int f_1n -(f_2+f_3)\cdot u dx
  +C\| E \|\| f_1\|_{L^{\frac65}}.  \label{basic.1}
 \ema
Since it is easy to verify
 \bma
 \|f_1\|_{L^{\frac65}}
 &\leq
  C\|(nDu+uDn)\|_{L^{\frac65}}
 \leq
  C\|n\|_{L^{3}}\|Du\|_{L^{2}}+C\|Dn\|_{L^{2}}\|u\|_{L^{3}}
  \nnm\\
 &\leq
   C(\|n\|_{L^{2}}+\|n\|_{L^{6}})\|Du\|_{L^{2}}
  +C\|Dn\|_{L^{2}}(\|u\|_{L^{2}}+\|u\|_{L^{6}})
 \nnm\\
 &\leq
  C(\|n\|+\|Dn\|)\|Du\|+C\|Dn\| \|u\|, \label{basic.3}
 \ema
we can control the last term on the right hand side of \ef{basic.1}
as
 \bma
\| E \|\| f_1\|_{L^{\frac65}}
&\leq
 C\| E\|(\|n\|^2+\|Dn\|^2+\|Du\|^2)+C\|Dn\|(\|u\|^2+\| E\|^2)
 \nnm\\
&\leq
 C\| E \|\|(n,Dn,Du)\|^2+C\|Dn\|\|(u,E)\|^2.\label{basic.4}
 \ema
Substituting \ef{basic.4} into \ef{basic.1} we have
  \bma
 &\frac12\frac{d}{dt}\int(|n|^2+|u|^2+|E|^2)(t)dx+C\|Du(t)\|^2dx
\nnm\\
 \leq&
  C\| E(t) \|\|(n,Dn,Du)(t)\|^2+C\|Dn\|\|(u,E)(t)\|^2
 +\int f_1n -(f_2+f_3)\cdot u dx
\nnm\\
\leq &
   C\Lambda_1(t)\|(n,Dn,Du)(t)\|^2
  +C(1+t)^{-\frac54}\Lambda_1(t)\|(u,E)(t)\|^2\nnm\\
 & +\int f_1n -(f_2+f_3)\cdot u dx.        \label{basic.5}
  \ema
Due to the facts
$$
 f_1=-n\div u-\nabla n\cdot u,\quad
 f_2\sim \mathcal{O}(1)(nDn+uDu+nD^2u),\quad
 f_3=nE,
$$
it is easy to estimate the last term on the right hand side of
\ef{basic.5} as
 \bma
 &\int f_1n -(f_2+f_3)\cdot u dx\nnm\\
 \leq&
 C\varepsilon\|Du(t)\|^2+C\Lambda_1(t)\|(n,Dn,Du)(t)\|^2
 +C\frac{1}{\varepsilon}\|u(t)\|_{L^{\infty}}^2\|u(t)\|^2\nnm\\
\leq&
 C(\varepsilon\|Du(t)\|^2+\Lambda_1(t)\|(n,Dn,Du)(t)\|^2
 +\frac{1}{\varepsilon}(1+t)^{-\frac32}\Lambda_1(t)\|u(t)\|^2),
 \ema
which together with \ef{basic.5} and the smallness of constant
$\varepsilon>0$ gives rise to
 \bma
 &\frac12\frac{d}{dt}\int(|n|^2+|u|^2+|E|^2)(t)dx+C\|Du(t)\|^2dx\nnm\\
\leq&
 C\Lambda_1(t)\|(n,Dn,Du)(t)\|^2+C(1+t)^{-\frac54}\Lambda_1(t)\|(u,E)(t)\|^2\nnm\\
& +C(1+t)^{-\frac32}\Lambda_1(t)\|u(t)\|^2.\label{basic.6}
  \ema

In a similar procedure, we are able to deal with the higher order
derivatives of the solution as follows
 \bma
 &\frac{d}{dt}\int(|D^kn|^2+|D^ku|^2+|D^kE|^2)(t)dx+C\|D^{k+1}u(t)\|^2dx\nnm\\
\leq&
 C\Lambda_1(t)(\|n(t)\|_{H^k}^2+\|Du(t)\|_{H^k}^2)\nnm\\
& +C(1+t)^{-\frac54}\Lambda_1(t)\|(u,E)(t)\|^2
 +C(1+t)^{-\frac32}\Lambda_1(t)\|u(t)\|^2,
 \label{basic.7}
  \ema
with $k=1,2,3,4,$ which implies
 \bma
 &\frac{d}{dt}(\|n(t)\|_{H^4}^2+\|u(t)(t)\|_{H^4}^2+\|E(t)(t)\|_{H^4}^2)
  +C\|Du(t)\|_{H^4}^2\nnm\\
\leq&
 C\Lambda_1(t)(\|n(t)\|_{H^4}^2+\|Du(t)\|_{H^4}^2)
\nnm\\
& +C(1+t)^{-\frac54}\Lambda_1(t)\|(u,E)(t)\|^2
  +C(1+t)^{-\frac32}\Lambda_1(t)\|u(t)\|^2.\label{basic.8}
  \ema

Taking inner product between \ef{pert1b'} and $\nabla n$,
integrating the resulted equation by parts over $\R^3$, and making
use of $\int E\cdot\nabla ndx=-\int\nabla\cdot Endx=\int|n|^2dx$, we
have after a straightforward computation that
 \bma
 &\frac{d}{dt}\int u\cdot\nabla ndx-\int u\nabla n_tdx+\int|\nabla
 n|^2+|n|^2dx\nnm\\
 &\leq
 (\frac12+\Lambda_1(t))\|(n,Dn)(t)\|^2+C(1+\Lambda_1(t))\|(Du,D^2u)(t)\|^2,
 \ema
which leads to
 \bma
 &\frac{d}{dt}\int u\cdot\nabla ndx+\int|\nabla
 n|^2+|n|^2dx\nnm\\
 &\leq C\int u\nabla n_tdx
 +C(1+\Lambda_1(t))\|(Du,D^2u)(t)\|^2\nnm\\
 &\leq C\int u\nabla (\nabla\cdot u+n\nabla\cdot u+u\nabla n)dx
 +C(1+\Lambda_1(t))\|(Du,D^2u)(t)\|^2\nnm\\
&\leq
C\Lambda_1(t)\|(n,Dn)(t)\|^2+C(1+\Lambda_1(t))\|Du(t)\|^2.\label{basic.9}
 \ema
Similarly, taking  $\int D^{\alpha}\ef{pert1b'}\cdot \nabla
D^{\alpha}ndx$ with $1\leq|\alpha|\leq3$, we can have after a
tedious and complicated calculation that
 \bma
 &\frac{d}{dt}\int D^{\alpha}u\cdot\nabla D^{\alpha} ndx+\int|\nabla
 D^{\alpha}n|^2+|D^{\alpha}n|^2dx\nnm\\
\leq
 &C\Lambda_1(t)\|n(t)\|_{H^4}^2+C(1+\Lambda_1(t))\|Du(t)\|_{H^3}^2,
\label{basic.10}
 \ema
which together with \ef{basic.9} and the a-priori smallness
assumption of $\Lambda_1(t)$ gives rise to
 \be
 \frac{d}{dt}\int D^{\alpha}u\cdot\nabla D^{\alpha} ndx+\|n(t)\|_{H^4}^2
 \leq
 C\|Du(t)\|_{H^3}^2,\quad 0\leq|\alpha|\leq3. \label{basic.11}
 \ee

Taking the summation  $\ef{basic.8}+\beta\times\ef{basic.11}$ with
$\beta>0$ small enough, and with $\Lambda(t)$ rather small we have
 \bma
  &
\frac{d}{dt}K(t)+C(\|n(t)\|_{H^4}^2+\|Du(t)\|_{H^4}^2)\nnm\\
\leq &
    C(1+t)^{-\frac54}\Lambda_1(t)\|(u,E)(t)\|^2
    +C(1+t)^{-\frac32}\Lambda_1(t)\|u(t)\|^2\nnm\\
\leq &
    C(1+t)^{-\frac54}\Lambda_1(t)K(t),
\label{basic.12}
  \ema
where
$$K(t)=(\|n(t)\|_{H^4}^2+\|u(t)(t)\|_{H^4}^2+\|E(t)(t)\|_{H^4}^2
         +\beta\int \sum_{|\alpha|\leq3}D^{\alpha}u\cdot\nabla D^{\alpha} ndx)
$$
and
 \be
 C(\|n(t)\|_{H^4}^2+\|u(t)\|_{H^4}^2
  +\|E(t)\|_{H^4}^2)\leq K(t)\leq C'(\|n(t)\|_{H^4}^2+\|u(t)\|_{H^4}^2
  +\|E(t)\|_{H^4}^2).\nnm
 \ee
From \ef{basic.12} we have by the Gronwall's inequality that
 \bma
 K(t)\leq
  Ce^{\int_0^t(1+\tau)^{-\frac54}\Lambda(\tau)d\tau}\|(n_0,u_0,E_0)\|_{H^4}
\leq C\delta.
 \ema
This together with \ef{basic.12} also leads to
 \be
\int_0^t\|(n,Du)(\tau)\|_{H^4}^2d\tau\leq C\delta,
 \ee
and finally
 \be
 \|(n,m,E)(t)\|_{H^4}\leq C\delta. \label{basic.13}
 \ee
\par

\underline{\it Step 3: Closure of the estimates \ef{claim 1}}.\ \
The combination of \ef{n L2 0}, \ef{n L2 1}--
\ef{E L2 0} and \ef{basic.13} leads to
 \be
 \Lambda_{1}(t)\leq C\delta+C\delta\Lambda_{1}(t)+C(\Lambda_{1}(t))^2,
 \quad t\in[0,T],
 \ee
which together with the smallness of $\delta>0$ leads to the first
estimate in \ef{claim 1}. The proof is completed.
 \hspace{\stretch{1}}$\square$

\bigskip

\underline{\textit{The proof of Theorem \ref{existence}  and
Theorem~\ref{Lower-bound}}}:\,  The global existence of smooth
solution of the IVP problem for the compressible
Navier-Stokes-Poisson system \ef{1.1a}--\ef{1.1d} follows from the
short time existence theory, the uniformly a-priori estimates, and
the continuity argument. The time-decay rate in Theorem
\ref{existence} follows from the Lemma~\ref{xin-lemma}. The optimal
time-decay rate in Theorem~\ref{Lower-bound} follows from the
combination of the Proposition~\ref{linear_theory_L2-a lower}, and
the uniform estimates \ef{claim 1}, \ef{n L2 0}, \ef{m L2 0}, and
\ef{E L2 0}.

\section{$L^p$- time decay rate}
\label{Lp-theory} \setcounter{equation}{0}
\subsection{$L^p$ decay rate for linear semigroup}
In this section, we investigate the $L^p$- time decay rate for the
solution of linearized NSP system, with $p\in[2,\infty]$. To this
end, we need to analyze the Green's function  $G=e^{tB}$ that formed
by the linearized NSP system.
 \par

We have the $L^p$ time decay rate for the linear semigroup as
follows.

 \oplem{\textbf{\boldmath ($L^p$ decay rate)}}\label{lp}
Let $(n,m)=G* U_0$ and $E=L\ast U_0$ with $N,M,L$  defined by
\eqref{define} and \eqref{define4}, then we have
 \bma
 \|D_x^\alpha n(t)\|_{L^p}=\|D_x^\alpha(N*U_0)(t)\|_{L^p}
  &\leq C(1+t)^{-\frac32(1-\frac1p)-\frac{|\alpha|}2}
        (\|U_0\|_{L^1}+\|D_x^{\alpha}U_0\|_{L^p}),\nnm\\
\|D_x^\alpha m(t)\|_{L^p}=\|D_x^\alpha(M*U_0)(t)\|_{L^p}
  &\leq C(1+t)^{-\frac32(1-\frac1p)+\frac12-\frac{|\alpha|}2}
        (\|U_0\|_{L^1}+\|D_x^{\alpha}U_0\|_{L^p}),\nnm\\
\|D_x^\alpha E(t)\|_{L^p}=\|D_x^\alpha(L*U_0)(t)\|_{L^p}
  &\leq C(1+t)^{-\frac32(1-\frac1p)+\frac12-\frac{|\alpha|}2}
        (\|U_0\|_{L^1}+\|D_x^{\alpha}U_0\|_{L^p}),\nnm
 \ema
where $p\in[2,\infty]$, and the constant $C>0$ independent of time.
\cllem

To prove Lemma~\ref{lp}, we shall make use of following two lemmas
\cite{hoff}.

\oplem{\textbf{\boldmath ($L^p$ multiplier,\cite{hoff})}}\label{lp
multiplier} Let $n\geq1$ and assume that $\widehat{f}(\xi)\in
L^\infty\cap C^{n+1}(R^n/\{0\}),$ with
 \bma
 |D_{\xi}^\alpha\widehat{f}(\xi)|
 \leq C^{'}\left\{\begin{aligned}
  &|\xi|^{-|\alpha|+\sigma_1}&\ \ \ &|\xi|\leq R;\ |\alpha|=n,\\
&|\xi|^{-|\alpha|-\sigma_2}&\ \ \ &|\xi|\geq R;\ |\alpha|=n-1,\ n,\
n+1,
\end{aligned}
\right.
\ema
 where $\sigma_1,\sigma_2>0$ and $n>2-2\sigma_2$. Then
 $\widehat{f}(\xi)$ is continuous at both 0 and $\infty$, and
$$f=m_1+m_2\delta,$$
where $m_1\in L^1(R^n)$ satisfies $\|m_1\|\leq C(C^{'})$, $m_2$
 is the constant
$$
 m_2=(2\pi)^{-\frac{n}{2}}\lim_{|\xi|\rightarrow\infty}\widehat{f}(\xi),
$$
and $\delta$ is the Dirac distribution. In particular,
$\widehat{f}(\xi)$ is a strong $L^p$ multiplier, $1\leq
p\leq\infty,$ in sense that, for any $g\in L^p$,
$$
\|f\ast g\|_{L^p}\leq C\|g\|_{L^p},\ \ 1\leq p\leq\infty,
$$
where $C>0$ depends   on $|m_2|\le \|\widehat{f}\|_{L^\infty}$ and
the constant $C^{'}$. \cllem

\oplem{\textbf{\boldmath(\cite{hoff})}}\label{lp result}
 Let $\widehat{g}(\xi,t)=\widehat{K}(\xi,t)\widehat{f}(\xi)$ with
 $\widehat{K}(\xi,t)=e^{-c_5|\xi|^2t}$ and
 $\widehat{f}(\xi)\in L^\infty\cap
 C^{n+1}(\R^n)$ satisfying
 \be
|D_{\xi}^{\beta}\widehat{f}(\xi)|\leq C^{'}|\xi|^{-|\beta|},\ \
|\beta|\leq n+1.
 \label{condition}\ee
Then,  for any $t>0$ $D_x^{\alpha}g(\cdot,t)\in L^p(\R^n)$ for all
$\alpha$ and $p\in[1,\infty]$, and it holds \be
 \|D_x^{\alpha}g(\cdot,t)\|_{L^p(\R^n)}
 \leq
 C(|\alpha|)t^{\frac{n}{2}(\frac1p-1)-\frac{\alpha}{2}}.
 \label{conclusion}
\ee
 \cllem

\bigskip

\underline{\textit{The proof of Lemma \ref{lp}} }: \, To make use of
the above lemmas to prove Lemma \ref{lp}, we decompose the Fourier
transform $\widehat{N},\widehat{M}, \widehat{L}$ into low frequency
term and high frequency term below.
\par

 Define
 \be\left\{\bln
 &\widehat{\mathcal{N}}:=\widehat{\mathcal{N}_1}+\widehat{\mathcal{N}_2}, \quad
\widehat{\mathfrak{N}}:=\widehat{\mathfrak{N}_1}+\widehat{\mathfrak{N}_2};\quad
\widehat{\mathcal{M}}:=\widehat{\mathcal{M}_1}+\widehat{\mathcal{M}_2}, \\
&\widehat{\mathfrak{M}}:=\widehat{\mathfrak{M}_1}+\widehat{\mathfrak{M}_2};\quad
 \widehat{\mathcal{L}}:=\widehat{\mathcal{L}_1}+\widehat{\mathcal{L}_2},\quad
 \widehat{\mathfrak{L}}:=\widehat{\mathfrak{L}_1}+\widehat{\mathfrak{L}_2},
 \label{decompose}
 \eln\right.
 \ee
  where
$(\cdot)_1=\chi(\xi)(\cdot),\ (\cdot)_2=(1-\chi(\xi))(\cdot)$, and
$\chi(\xi)$ is the smooth cut off function that
 \bma \chi(\xi)
 = \left\{\begin{aligned}
  &1&\ \ \ &|\xi|\leq R,\\
&0&\ \ \ &|\xi|\geq R+1.
\end{aligned}
\right.
 \ema
Thus, in terms of \ef{define} and \ef{decompose}, we have the
following decomposition of $(n,m)=G* U_0$ in Fourier modes
 \bma
 \widehat{n}
 =&\widehat{N}\cdot\widehat{U}_0
 =\widehat{N}_1\cdot\widehat{U}_0+\widehat{N}_2\cdot\widehat{U}_0
 =(\widehat{\mathcal{N}}_1+\widehat{\mathfrak{N}}_1)
  \cdot\widehat{U}_0
  +(\widehat{\mathcal{N}}_2+\widehat{\mathfrak{N}}_2)
  \cdot\widehat{U}_0,       \label{define-a}
\\
 \widehat{m}
 =&\widehat{M}_1\cdot\widehat{U}_0+\widehat{M}_2\cdot\widehat{U}_0
 =(\widehat{\mathcal{M}}_1+\widehat{\mathfrak{M}}_1)
  \cdot\widehat{U}_0
  +(\widehat{\mathcal{M}}_2+\widehat{\mathfrak{M}}_2)
   \cdot\widehat{U}_0. \label{define-b}
 \ema

Let we first analyze above  higher frequency terms denoted by
$\widehat{(\cdot)}_2$. Recall that
\bma
 \lambda_0=&-\mu|\xi|^2,\quad\label{bs0}
\\[2mm]
 \lambda_+
        =&-(\mu+\mbox{$\frac12$}\nu)|\xi|^2
            +\mbox{$\frac12$}i\sqrt{
             4(c^2|\xi|^{2}+ \lambda^{-2})
            -(2\mu +\nu)^2|\xi|^4}\nnm\\
        =&-(2\mu +\nu)|\xi|^2
            +\frac{c^2}{2\mu +\nu}+\mathcal{O}(|\xi|^{-1}),
            \quad|\xi|\gg1,\label{bs+}
\\[2mm]
 \lambda_-=&-(\mu+\mbox{$\frac12$}\nu)|\xi|^2
            -\mbox{$\frac12$}i\sqrt{
             4(c^2|\xi|^{2}+ \lambda^{-2})
            -(2\mu +\nu)^2|\xi|^4}\nnm\\
          =&-\mbox{$\frac{c^2}{2\mu +\nu}$}+\mathcal{O}(|\xi|^{-1}),
          \quad|\xi|\gg1.\label{bs-}
\ema
%
%
%
We shall prove  that the higher frequency terms are $L^p$ Fourier
multipliers with an exponential time decay coefficient $Ce^{-c_5t}$.
For simplicity, we only show that $\widehat{\mathcal{N}_2}$ is an
$L^p$ Fourier multiplier at higher frequency as follows. It holds
 \bma
 \mbox{$\frac{\lambda_+e^{\lambda_-t}-\lambda_-e^{\lambda_+t}}
         {\lambda_+-\lambda_-}$}
         = \mbox{$\frac{\lambda_+e^{\lambda_-t}-\lambda_-e^{\lambda_-t}
            +(\lambda_-e^{\lambda_-t}-\lambda_-e^{\lambda_+t})}
         {\lambda_+-\lambda_-}$}
        =e^{\lambda_-t}
           +\mbox{$\frac{\lambda_-e^{\lambda_-t}}{\lambda_+-\lambda_-}
           -\frac{\lambda_-e^{\lambda_+t}}{\lambda_+-\lambda_-}$},
     \nnm
 \ema
and
 \bma
 \lambda_{-}(\xi)
 =&-(\mu+\frac\nu2)|\xi|^2-\frac{i}2
                  \sqrt{4(c^2|\xi|^2+\lambda^{-2})-(2\mu+\nu)^2|\xi|^4}\nnm\\
  =&\mbox{$
    -\frac{c^2}{2\mu+\nu}
    +\{\frac{c^2}{2\mu+\nu}-(\mu+\frac\nu2)|\xi|^2\}
    +\frac{1}2\sqrt{(2\mu+\nu)^2|\xi|^4-4(c^2|\xi|^2+\lambda^{-2})}
    $}
  \nnm\\
 =&\mbox{$
    -\frac{c^2}{2\mu+\nu}
    +\frac{\lambda^{-2}+(\frac{c^2}{2\mu+\nu})^2}
          {\frac{1}2\sqrt{(2\mu+\nu)^2|\xi|^4-4(c^2|\xi|^2+\lambda^{-2})}
            +(\mu+\frac\nu2)|\xi|^2-\frac{c^2}{2\mu+\nu}},
     $}\nnm
 \ema
where
 \be
\sqrt{(2\mu+\nu)^2|\xi|^4-4(c^2|\xi|^2+\lambda^{-2})}\sim
\mathcal{O}(|\xi|^2),\quad\quad|\xi|\gg1.
\ee
By a direct
computation, it is easy to verify
 \be
|D_{\xi}^{\alpha}\lambda_{-}|\leq
C|\xi|^{2+\alpha},\quad\quad|\xi|\gg1,\label{qd}
 \ee
which together with \ef{bs+}--\ef{bs-} gives rise to
 \bma
|D_{\xi}^{\alpha}[(1-\chi(\cdot))e^{\lambda_{-}t}]|
 \leq C\left\{\begin{aligned}
  &0&&|\xi|\leq R,\\
&e^{-c_6t}|\xi|^{-|\alpha|}|\xi|^{-2}&&|\xi|\geq R,
\end{aligned}
\right.
 \ema
where and below $R>0$ is a given constant,  and
 \bma
|D_{\xi}^{\alpha}[(1-\chi(\cdot))
   \mbox{$ \frac{\lambda_-e^{\lambda_-t}}{\lambda_+-\lambda_-}$}]|
   \leq C
   \left\{\begin{aligned}
  &0&&|\xi|\leq R,\\
  &e^{-c_6t}|\xi|^{-|\alpha|}|\xi|^{-2}&&|\xi|\geq R.
        \end{aligned}\right.
  \ema
Thus, from Lemma \ref{lp multiplier} it follows that the inverse
Fourier transform of the term $
 (1-\chi(\cdot))
 (e^{\lambda_{-}t}
  +\frac{\lambda_-e^{\lambda_-t}}{\lambda_+-\lambda_-})
$  
is an $L^p$ multiplier with the coefficient  $Ce^{-c_6t}$. The other
part of $\widehat{\mathcal{N}_2}$ at higher frequency can be written
as
 \be
\mbox{$
(1-\chi(\cdot))\frac{\lambda_-e^{\lambda_+t}}{\lambda_+-\lambda_-}
 =e^{-(\mu+\frac\nu2)|\xi|^2t}\cdot
   (\frac{e^{(\lambda_+-(\mu+\frac\nu2)|\xi|^2)t}}{\lambda_+-\lambda_-}
    (1-\chi(\cdot)))$},\nnm
 \ee
whose first term in right hand side can be looked on as the function
$K(\xi,t)$ of Lemma \ref{lp result}, and the rest term satisfies the
condition \ef{condition} due to the facts
$(\lambda_+-(\mu+\frac\nu2)|\xi|^2)t\sim -c_6|\xi|^2t$ for
$|\xi|\geq1$ and $D^k(e^{-|\xi|^2t})\leq
C|\xi|^{-k}((|\xi|^2t)+(|\xi|^2t)^2+\cdots+(|\xi|^2t)^k)e^{-|\xi|^2t}\leq
C|\xi|^{-k}$. Thus, the inverse Fourier transform of above term is
also an $L^p$ multiplier with the coefficient $Ce^{-c_6R^2t}$. These
facts imply that $\mathcal{N}_2$ is an $L^p$ multiplier with the
coefficient $Ce^{-c_7t}$.

Applying the similar analysis to the terms
$\widehat{\mathfrak{N}_2}$, $\widehat{\mathcal{M}_2}$,
$\widehat{\mathfrak{M}_2}$, $\widehat{\mathcal{L}_2}$, and
$\widehat{\mathfrak{L}_2}$, we can show that their inverse Fourier
transform are all $L^p$ multipliers with the constant $Ce^{-c_8t}$.
Thus taking $c_5=\min\{c_7,c_8\}$ and then
 \be
 \|D^\alpha_x(N_2*f),D^\alpha_x(M_2*f),D^\alpha_x(L_2*f)\|_{L^p}
 \leq Ce^{-c_5t}\|D^\alpha f\|_{L^p}\label{DH},\ee
for all $|\alpha|\ge 0$, and $p\in[2,\infty]$.
 \par
To prove Lemma \ref{lp}, we also need to deal with the corresponding
lower frequency terms denoted by $\widehat{(\cdot)_1}$. Indeed, we
can have
 \bma
&\|D_x^\alpha \mathcal{N}_1(t)\|_{L^p}
  \leq C(1+t)^{-\frac32(1-\frac1p)-\frac{|\alpha|}2},\quad
   &&\|D_x^\alpha \mathfrak{N}_1(t)\|_{L^p}
  \leq C(1+t)^{-\frac32(1-\frac1p)-\frac12-\frac{|\alpha|}2},
  \label{ts1}\\
 &\|D_x^\alpha \mathfrak{M}_1(t)\|_{L^p}
  \leq C(1+t)^{-\frac32(1-\frac1p)+\frac12-\frac{|\alpha|}2},\quad
   &&\|D_x^\alpha \mathcal{M}_1(t)\|_{L^p}
  \leq C(1+t)^{-\frac32(1-\frac1p)-\frac{|\alpha|}2},
  \label{ts2}\\
  &\|D_x^\alpha \mathcal{L}_1(t)\|_{L^p}
  \leq C(1+t)^{-\frac32(1-\frac1p)+\frac12-\frac{|\alpha|}2},\quad
   &&\|D_x^\alpha \mathfrak{L}_1(t)\|_{L^p}
  \leq C(1+t)^{-\frac32(1-\frac1p)-\frac{|\alpha|}2},
  \label{ts3}
  \ema
for $t\geq0 $ and $p\in [2,\infty]$.
\par

For simplicity, we only prove \ef{ts1} and \ef{ts2}, since the
analysis of \ef{ts3} can be carried out in a similar way. By
\ef{xs1}--\ef{xs3}, we have the asymptotical approximation
 \bgr
 \mbox{$  \frac{\lambda_+e^{\lambda_-t}-\lambda_-e^{\lambda_+t}}
         {\lambda_+-\lambda_-}\sim \
   \mathcal{O}(1)e^{-(\mu+\frac12\nu)|\xi|^2t},\quad
         \frac{\lambda_+e^{\lambda_+t}-\lambda_-e^{\lambda_-t}}
         {\lambda_+-\lambda_-}\sim \
   \mathcal{O}(1)e^{-(\mu+\frac12\nu)|\xi|^2t}$},\quad
  |\xi|\ll1,\nnm
   \\
        \mbox{$  \   \frac{e^{\lambda_+t}- e^{\lambda_-t}}
         {\lambda_+-\lambda_-}
\sim \
   \mathcal{O}(1)e^{-(\mu+\frac12\nu)|\xi|^2t},\quad
  |\xi|\ll1,$}\nnm
 \egr
which imply for  $|\xi|\ll1$ that
 \be
 \left\{\begin{aligned}
 &|\widehat{\mathcal{N}_1}|\sim \mathcal{O}(e^{-c_9|\xi|^2t}), \quad
|\widehat{\mathfrak{N}_1}|\sim \mathcal{O}(e^{-c_9|\xi|^2t}|\xi|),
\\
 &|\widehat{\mathfrak{M}_1}|\sim
 \mathcal{O}(e^{-c_9|\xi|^2t}\frac1{|\xi|}),\quad
|\widehat{\mathcal{M}_1}|\sim
 \mathcal{O}(1)e^{-c_9|\xi|^2t}.
 \label{dpjs}
 \end{aligned}\right.
 \ee
Thus, by Hausdroff-Young's inequality, we can have
\bgr
\|D_x^\alpha\mathcal{N}_1(t)\|_{L^p}
\leq
 C\{\int_{|\xi|\le \eta}||\xi|^{|\alpha|}e^{-c_9|\xi|^2t}|^qd\xi\}^{\frac1q}
 \leq
 Ct^{-\frac32(1-\frac1p)-\frac{|\alpha|}{2}},\label{js1}
\\
 \|D_x^\alpha\mathfrak{N}_1(t)\|_{L^p}
 \leq
 C\{\int_{|\xi|\le \eta}||\xi|^{|\alpha|+1}e^{-c_9|\xi|^2t}|^qd\xi\}
     ^{\frac1q}
 \leq
  Ct^{-\frac32(1-\frac1p)-\frac12-\frac{|\alpha|}{2}},
 \label{js2}
\egr with $\frac1p+\frac1q=1$ and $p\in[2,\infty]$, and
 \bgr
 \|D_x^\alpha \mathfrak{M}_1(t)\|_{L^p}
\leq
 C\{\int_{|\xi|\le \eta}||\xi|^{|\alpha|-1}e^{-c_9|\xi|^2t}|^qd\xi\}^{\frac1q}
  \leq C(1+t)^{-\frac32(1-\frac1p)+\frac12-\frac{|\alpha|}2},
 \\
 \|D_x^\alpha \mathcal{M}_1(t)\|_{L^p}
 \leq
 C\{\int_{|\xi|\le \eta}||\xi|^{|\alpha|}e^{-c_9|\xi|^2t}|^qd\xi\}^{\frac1q}
  \leq
  C(1+t)^{-\frac32(1-\frac1p)-\frac{|\alpha|}2}.
  \label{ts2-b}
 \egr
These give  rise to \ef{ts1} and  \ef{ts2}.
\par

Combining the estimates \ef{ts1}--\ef{ts3} 
 and \ef{DH}, 
we finally have for $t>0$ that
 \bma
 \|D_x^\alpha(N*f)(t)\|_{L^p}
 =&\|D_x^\alpha((N_1+N_2)*f)\|_{L^p}
\nnm\\
\leq
 &C(1+t)^{-\frac32(1-\frac1p)-\frac{|\alpha|}{2}}\|f\|_{L^1}
     +Ce^{-c_6t}\|D_x^\alpha f\|_{L^p}
\nnm\\
 \leq&
 C(1+t)^{-\frac32(1-\frac1p)-\frac{|\alpha|}{2}}
     (\|f\|_{L^1}+\|D_x^\alpha f\|_{L^p}),
 \quad\quad p\in[2,\infty],\nnm
\ema where ${N_i}={\mathcal{N}_i}+{\mathfrak{N}_i}$, $i=1,2$, and
similarly
 \bgr
 \|D_x^\alpha(M*f)(t)\|_{L^p}
 \leq C(1+t)^{-\frac32(1-\frac1p)+\frac12-\frac{|\alpha|}{2}}
     (\|f\|_{L^1}+\|D_x^\alpha f\|_{L^p}),\quad p\in[2,\infty],\nnm
\\
\|D_x^\alpha(L*f)(t)\|_{L^p}\leq
C(1+t)^{-\frac32(1-\frac1p)+\frac12-\frac{|\alpha|}{2}}
     (\|f\|_{L^1}+\|D_x^\alpha f\|_{L^p}),\quad p\in[2,\infty].\nnm
\egr
The proof of Lemma \ref{lp} is
completed.\hspace{\stretch{1}}$\square$
\par

In addition to Lemma \ref{lp}, we can also have following estimates,
which can be shown in a similar procedure as in the proof of Lemma
\ref{lp} and are applicable to the $L^p$ time decay rate of solution
of the original IVP problem for the nonlinear Navier-Stokes-Poisson
system. Indeed, we have

\oplem{\textbf{}}\label{lpf} It holds for the Green's functions
$\mathfrak{N}$, $\mathcal{M}$, and $\mathfrak{L}$ defined by
\eqref{define1}--\eqref{define3} that
 \bma
 \|D_x^\alpha (\mathfrak{N}*f)(t)\|_{L^p}
  &\leq C(1+t)^{-\frac32(1-\frac1p)-\frac12-\frac{|\alpha|}2}
        (\|f\|_{L^1}+\|D_x^{\alpha}f\|_{L^p}),\label{yd1}\\
\|D_x^\alpha (\mathfrak{N}*f)(t)\|_{L^p}
  &\leq C(1+t)^{-\frac32(\frac12-\frac1p)-\frac12-\frac{|\alpha|}2}
        (\|f\|_{L^2}+\|D_x^{\alpha}f\|_{L^p}),\label{yd2}\\
 \|D_x^\alpha (\mathcal{M}*f)(t)\|_{L^p}
  &\leq C(1+t)^{-\frac32(1-\frac1p)-\frac{|\alpha|}2}
        (\|f\|_{L^1}+\|D_x^{\alpha}f\|_{L^p}),\label{yd3}\\
     \|D_x^\alpha (\mathcal{M}*f)(t)\|_{L^p}
  &\leq C(1+t)^{-\frac32(\frac12-\frac1p)-\frac{|\alpha|}2}
        (\|f\|_{L^2}+\|D_x^{\alpha}f\|_{L^p}),\label{yd3'}\\
 \|D_x^\alpha (\mathfrak{L}*f)(t)\|_{L^p}
  &\leq C(1+t)^{-\frac32(1-\frac1p)-\frac{|\alpha|}2}
        (\|f\|_{L^1}+\|D_x^{\alpha}f\|_{L^p}),\label{yd4}\\
    \|D_x^\alpha (\mathfrak{L}*f)(t)\|_{L^p}
  &\leq C(1+t)^{-\frac32(\frac12-\frac1p)-\frac{|\alpha|}2}
        (\|f\|_{L^2}+\|D_x^{\alpha}f\|_{L^p}),\label{yd4'}
  \ema
with $p\in[2,\infty]$, $|\alpha|=k\ge 0$, and the constant $C>0$
independent of time, so long as $f\in L^2(\R^n)\cap W^{k,p}(\R^n)$
or  $f\in L^1(\R^n)\cap W^{k,p}(\R^n)$.
\cllem
\textbf{Proof:} The estimates \ef{yd1}, \ef{yd3} and \ef{yd4} follow
directly from  those of \ef{ts1}--\ef{ts3} and the related estimates
for higher frequency terms. We need only to prove \ef{yd2},
\ef{yd3'} and \ef{yd4'}. Indeed, it holds for $p\in(2,\infty]$ that
 \bma
\|D_x^\alpha (\mathfrak{N}*f)(t)\|_{L^p}
 \leq
 &\|D_x^\alpha(\mathfrak{N}_1*f)\|_{L^p}
  +\|D_x^\alpha (\mathfrak{N}_2*f)\|_{L^p}
\leq
   \|D_x^\alpha (\mathfrak{N}_1*f)\|_{L^p}
  + Ce^{-c_7t}\|D_x^\alpha f\|_{L^p}\nnm
 \ema
and
 \bma
\|D_x^\alpha (\mathfrak{N}_1*f)(t)\|_{L^p}
 \leq& C\|(i\xi)^\alpha\widehat{\mathfrak{N}_1}\hat f\|_{L^q}\nnm\\
 \leq&C\big[\big(\int_{|\xi|<\eta}
               (|\xi|^{|\alpha|+1}e^{-c_5|\xi|^2t})^{q\cdot\frac2{2-q}}d\xi
            \big)^{\frac{2-q}{2}}
              \cdot(\||\widehat{f}|^q\|_{L^{\frac2q}})
       \big]^{\frac1q}\nnm\\
\leq&
C\|f\|_{L^2}(1+t)^{-\frac32(\frac12-\frac1p)-\frac12-\frac{|\alpha|}{2}}
 \nnm
\ema
with $\frac1p+\frac1q=1$. As for $p=2$, a similar direct computation
gives rise to
\be
\|D_x^\alpha (\mathfrak{N}_1*f)(t)\|_{L^2}
 \leq C\|(i\xi)^\alpha\widehat{\mathfrak{N}_1}\hat f\|_{L^2}
 \leq C(1+t)^{-\frac12-\frac{|\alpha|}{2}}\cdot \|f\|_{L^2}.
\ee
Thus, we obtain the estimate \ef{yd2}. The proof of \ef{yd3'} and
\ef{yd4'} can be  completed in a similar fashion and the details are
omitted. \hspace{\stretch{1}}$\square$

\subsection{$L^p$ decay rate for nonlinear system}
In this subsection, we show the $L^p$ $(p\in[2,\infty])$ time decay
estimates for the original nonlinear problem. We can verify that the
expression of the solution of the IVP
problem~\eqref{pert1a}--\eqref{pert1d} is
 \bma
 &n=N\ast U_0+\int_0^t\mathfrak{N}(t-\tau)\ast Q(U)(\tau)d\tau,\label{express n'}
 \\
 &m=M\ast U_0+\int_0^t\mathcal{M}(t-\tau)\ast Q(U)(\tau)d\tau,\label{express m'}
 \\
 &E=L\ast U_0+\int_0^t\mathfrak{L}(t-\tau)\ast Q(U)(\tau)d\tau.\label{express E'}
 \ema

We have  Theorem  \ref{point-rate} concerned with the $L^p$ time
decay of strong solutions below.

\oplem{\textbf{}}\label{lpfq} Under the assumptions of Theorem~\ref{point-rate},
the global solutions $(n, m, E)$ with $E=\nabla\Phi$ of the IVP
problem~\eqref{pert1a}--\eqref{pert1d} satisfies
 \bma
 \|n(t)\|_{L^p}
  &\leq C(1+t)^{-\frac32(1-\frac1p)}
        (\|U_0\|_{L^1}+\|U_0\|_{L^p}),\label{yd1'}\\
\|(m,E)(t)\|_{L^p}
  &\leq C(1+t)^{-\frac32(1-\frac1p)+\frac12}
        (\|U_0\|_{L^1}+\|U_0\|_{L^p}),\label{yd1''}
  \ema
with $p\in[2,\infty]$, $U_0=(n_0,m_0)$, and the constant $C>0$
independent of time.
 \cllem

\textbf{Proof:} By Lemma \ref{xin-lemma}, we have for $l\geq5$ that
 \bma
&\|D^k n\|_{L^2}\leq C(1+t)^{-\frac34-\frac{k}{2}}\delta,\quad
k=0,1,2,&& \|D^3n\|_{L^2}\leq
C(1+t)^{-1}\delta,\nnm\\
&\|D^k (m, E)\|_{L^2}\leq C(1+t)^{-\frac14-\frac{k}{2}}\delta,\quad
k=0,1,2,&& \|D^3n\|_{L^2}\leq C(1+t)^{-1}\delta,\nnm
 \ema
where $\delta=: \|(\rho_0-\bar{\rho},m_0)\|_{H^l(\R^3)\cap
L^1(\R^3)}>0$ as defined in Theorem  \ref{point-rate}.
\par

Noticing that the nonlinear term $Q(U)=\nabla\cdot H$ contains
mainly those like $\mathcal{O}\{mD^2n, Dm\cdot Dn, nD^2m, mDm$,
$nDn, nE \}$, and using Sobolev embedding theorem, we can obtain
 \be
 \|Q(U)\|_{L^1}\leq C\delta^2(1+t)^{-1}, \|Q(U)\|_{L^p}
 \leq
C\delta^2(1+t)^{-\frac74},\quad p\in[2,\infty].   \label{zq1}
 \ee
Thus, by \ef{express n'}, \ef{zq1},  Lemma~\ref{lp} and
Lemma~\ref{lpf}, we are able  to show
 \bma
 \|n(t)\|_{L^p}
 \leq & \|N*U_0(t)\|_{L^p}+\int_0^t
       \|\mathfrak{N}(t-\tau)\ast Q(U)(\tau)\|_{L^p}d\tau
 \nnm\\
 \leq &C(1+t)^{-\frac32(1-\frac1p)}(\|U_0\|_{L^1}+\|U_0\|_{L^p})
  +C\int_0^t\|\mathcal N(t-\tau)*Q(U)(\tau)\|_{L^p}d\tau
\nnm\\
\leq &C(1+t)^{-\frac32(1-\frac1p)}(\|U_0\|_{L^1}+\|U_0\|_{L^p})
\nnm\\
     &+C\int_0^{\frac t2}(1+t-\tau)^{-\frac32(1-\frac1p)-\frac12}
      (\|Q(U)\|_{L^1}+\|Q(U)\|_{L^p})d\tau
\nnm\\
      &+C\int_{\frac t2}^t(1+t-\tau)^{-\frac32(\frac12-\frac1p)-\frac12}
      (\|Q(U)\|_{L^2}+\|Q(U)\|_{L^p})d\tau
\nnm\\
\leq
 & C(1+t)^{-\frac32(1-\frac1p)}(\|U_0\|_{L^1}+\|U_0\|_{L^p})
      +C(1+t)^{-\frac32(1-\frac1p)-\varepsilon}\delta^2
 \nnm\\
\leq& C(1+t)^{-\frac32(1-\frac1p)}\delta,\label{ok1}
\ema
and
\bma
\|m(t)\|_{L^p}
 \leq & \|M*U_0(t)\|_{L^p}+\int_0^t
       \|\mathcal{M}(t-\tau)\ast Q(U)(\tau)\|_{L^p}d\tau\nnm\\
 \leq &C(1+t)^{-\frac32(1-\frac1p)+\frac12}(\|U_0\|_{L^1}+\|U_0\|_{L^p})
  +C\int_0^t\|\mathcal M(t-\tau)*Q(U)(\tau)\|_{L^p}d\tau\nnm\\
  \leq &C(1+t)^{-\frac32(1-\frac1p)+\frac12}(\|U_0\|_{L^1}+\|U_0\|_{L^p})\nnm\\
     &+C\int_0^{\frac t2}(1+t-\tau)^{-\frac32(1-\frac1p)}
      (\|Q(U)\|_{L^1}+\|Q(U)\|_{L^p})d\tau
 \nnm\\
      &+C\int_{\frac t2}^t(1+t-\tau)^{-\frac32(\frac12-\frac1p)}
      (\|Q(U)\|_{L^2}+\|Q(U)\|_{L^p})d\tau
 \nnm\\
\leq
&C(1+t)^{-\frac32(1-\frac1p)+\frac12}(\|U_0\|_{L^1}+\|U_0\|_{L^p})
     +C\int_0^{\frac t2}(1+t-\tau)^{-\frac32(1-\frac1p)}
      (1+\tau)^{-1}d\tau
\nnm\\
      &+C\int_{\frac t2}^t(1+t-\tau)^{-\frac32(\frac12-\frac1p)}
      (1+\tau)^{-\frac74}d\tau
\nnm\\
\leq& C(1+t)^{-\frac32(1-\frac1p)+\frac12}\delta
+C(1+t)^{-\frac32(1-\frac1p)+\frac12-\varepsilon}\delta^2
\nnm\\
\leq& C(1+t)^{-\frac32(1-\frac1p)+\frac12}\delta,\label{ok2}
 \ema
where  $p\in[2,\infty]$, $C>0$ is a constant independent of time,
and $\varepsilon>$ is a small but fixed constant.
\par

The estimates of $\|E\|_{L^p}$ can be obtained in terms of $\|m\|$
and Riesz potential as
\be
 \|E(t)\|_{L^p}
 \leq
  C(1+t)^{-\frac32(1-\frac1p)+\frac12}
  (\|U_0\|_{L^1}+\|U_0\|_{L^p}),  \quad p\in[2,\infty].
\ee
The proof of the Lemma~\ref{lpfq} is completed.
 \hspace{\stretch{1}}$\square$
\par

\bigskip

 \vspace{0.5cm}

\noindent\textbf{Acknowledgements:}\, The authors acknowledge the
partial support by the National Natural Science Foundation of China
(No.10571102, No.10431060, No.10871134), the Key Research Project on
Science and Technology of the Ministry of Education of China
(No.104072), the Beijing Nova program 2005B48, the NCET support of
the Ministry of Education of China, the Institute of Mathematics and
Interdisciplinary Science at CNU, and the Huo Ying Dong Foundation
No.111033.


\begin{thebibliography}{99}
\normalsize
  \renewcommand\baselinestretch{1.0}
  \parskip=-5pt

\bibitem{D1992} K. Deckelnick,
Decay estimates for the compressible Navier-Stokes equations in
unbounded domains, \emph{Math. Z}. \textbf{209} (1992) 115--130.

\bibitem{D1993} K. Deckelnick,
$L^2$-decay for the compressible Navier-Stokes equations in
unbounded domains, \emph{Comm. Partial Diff. Eqns}. \textbf{18}
(1993) 1445--1476.

\bibitem{DFP113-130} B. Ducomet, E. Feireisl, H. Petzeltova and
I. S. Skraba, Global in time weak solution for compressible
barotropic self-gravitating fluids. \emph{Discrete  Continous
Dynamical System}, \textbf{11}, No 1 (2004), 113--130.


\bibitem{D31-37} B. Ducomet, A remark about global existence for
the Navier-Stokes-Poisson system, \emph{Applied Mathematics Letters}
\textbf{12} (1999) 31--37.

\bibitem{DD345-361} D. Donatelli,
Local and global existence for the coupled Navier-Stokes-Poisson
problem, \emph{Quart. Appl. Math}. \textbf{61} (2003), no. 2,
345--361.

\bibitem{D2007} R.-J. Duan, S. Ukai, T. Yang, and H.-J.  Zhao,
Optimal convergence rates for the compressible navier-Stokes
equations with potential forces, \emph{Mathematical Models and
Methods in Applied Sciences}, \textbf{17} (2007), no. 5 737--758.

\bibitem{D2008} R.-J. Duan, H. Liu, S. Ukai and T. Yang,
Optimal $L^p$-$L^q$ convergence rates for the compressible
Navier每Stokes equations with potential force, \emph{J. Diff. Eqns}.
\textbf{238} (2007), no. 1, 220--233


\bibitem{hoff} D. Hoff and K. Zumbrun, Multi-dimensional diffusion waves for
the Navier-Stokes equations of compressible flow, \emph{Indiana
University Mathematics Journal}, \textbf{44} (1995), 603--676.

\bibitem{H1997} D. Hoff and K. Zumbrun,
Pointwise decay estimates for multidimensional Navier-Stokes
diffusion waves, \emph{Z. Angew. Math. Phys}. \textbf{48} (1997)
597--614.





\bibitem{KK2002}  Y. Kagei, T. Kobayashi, On large time behavior of solutions to
the compressible Navier每Stokes equations in the half space in
$R^3$, \emph{Arch. Ration. Mech. Anal.} \textbf{165} (2002) 89每159.

\bibitem{KK2005} Y. Kagei, T. Kobayashi, Asymptotic behavior of solutions of the
compressible Navier-Stokes equations on the half space, \emph{Arch.
Ration. Mech. Anal.} \textbf{177} (2005) 231每330.

\bibitem{K2006} Y. Kagei and S. Kawashima,  Stability of planar stationary
solutions to the compressible Navier-Stokes equation on the half
space. \emph{Comm. Math. Phys}. \textbf{266} (2006), 401--430.

\bibitem{kawashima1987} S. Kawashima, Large-time behavior of solutions
to hyperbolic-parabolic systems of conservation laws and
applications, \emph{Proc. Roy. Soc. Edinburgh}, \textbf{106A}
(1987), 169--194.

\bibitem{K2003} S. Kawashima, S. Nishibata and P. Zhu,
Asymptotic stability of the stationary solution to the compressible
Navier-Stokes equations in the half space, \emph{Comm. Math. Phys},
\textbf{240} (2003), no. \textbf{3}, 483--500.

\bibitem{Ko2002}  T. Kobayashi, Some estimates of solutions for the equations of
motion of compressible viscous fluid in an exterior domain in $R^3$,
\emph{J. Differential Equations}, \textbf{184} (2002) 587每619.

\bibitem{KS1999} T. Kobayashi and Y. Shibata,
Decay estimates of solutions for the equations of motion of
compressible viscous and heat-conductive gases in an exterior domain
in $R^3$, \emph{Comm. Math. Phys}. \textbf{200} (1999), 621每659.

\bibitem{LW1998} T.-P. Liu and W.-K. Wang, The pointwise
estimates of diffusion waves for the Navier-Stokes equations in odd
multi-dimensions, \emph{Comm. Math. Phys}. \textbf{196} (1998),
145--173.

\bibitem{liu taiping} T.-P. Liu and Y. Zeng, Large time behavior
of solutions of general quasilinear hyperbolic-parabolic
 systems of conservation laws, \emph{A. M. S. Memoirs}
 \textbf{599} (1997), 1--45.

\bibitem{D.L.Li2005} D. L. Li, The Green's function of the Navier-Stokes
equations for gas dynamics in $R^3$, \emph{Comm. Math. Phys},
\textbf{257} (2005), 579--619.

\bibitem{matsumura1979} A. Matsumura and T. Nishida, The initial value
problem for the equation of motion of compressible viscous and
heat-conductive fluids, \emph{Proc. Japan. Acad.}, \textbf{55}
(1979), Ser. A, 337--342.

\bibitem{matsumura1980} A. Matsumura and T. Nishida, The initial value
problem for the equation of motion of viscous and heat-conductive
gases, \emph{J. Math. Kyoto. Univ.}, \textbf{20} (1980), 67--104.

\bibitem{matsumura1983} A. Matsumura and T. Nishida, Initial
boundary value problems for the equations of motion of compressible
viscous and heat conductive fluids, \emph{Comm. Math. Phys.},
\textbf{89} (1983), no.2, 445-464.

\bibitem{MRS1990}P. A.  Markowich, C. A. Ringhofer and C. Schmeiser,
 \emph{Semiconductor Equations}, Springer, 1990.

\bibitem{P1985} G. Ponce, Global existence of small solution to a class of
nonlinear evolution equations, \emph{Nonl. Anal.} \textbf{9} (1985)
339每418.

\bibitem{S257-275} V. A. Solonnikov,
Evolution free boundary problem for equations of motion of viscous
compressible selfgravitating fluid, \emph{SAACM} \textbf{3} (1993),
257--275.

\bibitem{ST2003} Y. Shibata, K. Tanaka,
On the steady compressible viscous fluid and its stability with
respect to initial disturbance, \emph{J. Math. Soc. Japan},
\textbf{55} (2003), 797-826.

\bibitem{STpreprint} Y. Shibata and K. Tanaka,
Rate of convergence of non-stationary flow to the steady flow of
compressible viscous fluid, preprint.

\bibitem{Uhuijiang} S. Ukai, T. Yang, and H.-J. Zhao,
Convergence rate for the compressible Navier每Stokes equations with
external force, \emph{J. Hyperbolic Diff. Eqns.}, \textbf{3} (2006),
561每574.

\bibitem{zeng} Y. Zeng, $L^1$ Asymptotic behavior of compressible isentropic
viscous 1-D flow. \emph{Comm. Pure Appl. Math.} \textbf{47} (1994),
1053每1082.


\bibitem{ZZ305-329} Y.-H. Zhang and Z. Tan,
On the existence of solutions to the Navier-Stokes-Poisson equations
of a two-dimensional compressible flow. \emph{Math. Meth. Appl.
Sci}, \textbf{30} (2007), 305--329.




\end{thebibliography}
\end{document}